\documentclass[12pt, draftclsnofoot, onecolumn]{IEEEtran}
\usepackage{amsmath,amsfonts}
\usepackage{color}
\usepackage{algorithmic}
\usepackage{algorithm}
\usepackage{array}
\usepackage{textcomp}
\usepackage{stfloats}
\usepackage{url}
\usepackage{verbatim}
\usepackage{graphicx}
\usepackage{cite}
\usepackage{subfigure,multirow,bm,stfloats}
\usepackage{float}
\begin{document}

\title{Radiomap Inpainting for Restricted Areas based on Propagation Priority and Depth Map}

\author{Songyang Zhang, Tianhang Yu, Brian Choi~\IEEEmembership{Senior Member,~IEEE,} Feng Ouyang~\IEEEmembership{Senior Member,~IEEE,} Zhi Ding~\IEEEmembership{Fellow,~IEEE,}
\thanks{This work was supported in part by the National Science Foundation under Grant 2029848 and  Grant 2002937; and in part by Johns Hopkins University Applied Physics Laboratory Independent Research and Development Funds.}
\thanks{S. Zhang, T. Yu and Z. Ding are with Department of Electrical and Computer Engineering, University of California, Davis, CA, 95616. (E-mail: \{sydzhang, thgyu, zding\}@ucdavis.edu,).}
\thanks{B. Choi and F, Ouyang are with Johns Hopkins University Applied Physics Laboratory, Laurel, Maryland, USA, 20723. (E-mail: \{Brian.Choi, Feng.Ouyang\}@jhuapl.edu)}
}

\markboth{Journal of \LaTeX\ Class Files,~Vol.~14, No.~8, August~2021}%
{Shell \MakeLowercase{\textit{et al.}}: A Sample Article Using IEEEtran.cls for IEEE Journals}


\maketitle

\begin{abstract}
Providing rich and useful information
regarding spectrum activities and propagation channels, radiomaps characterize the detailed distribution of  power spectral density (PSD) and are important tools for network planning in modern wireless systems.
Generally, radiomaps are constructed from radio strength measurements by
deployed sensors and user devices.
However, not all areas are accessible for radio
measurements due to physical constraints and security consideration, 
leading to non-uniformly spaced measurements and blanks on a radiomap.
In this work, we explore  distribution of radio spectrum strengths {in view of surrounding environments,} and propose two radiomap inpainting approaches for the 
reconstruction of radiomaps 
that cover missing areas. Specifically, we first define a propagation-based priority and integrate exemplar-based inpainting with radio propagation model for fine-resolution small-size missing area reconstruction on a radiomap.
Then, we introduce a novel radio depth map and propose a two-step template-perturbation approach for large-size restricted region inpainting. Our experimental results demonstrate the power of the proposed propagation priority and radio depth map in capturing the PSD distribution, as well as the efficacy of the proposed methods for radiomap reconstruction.
\end{abstract}


\section{Introduction}
\begin{figure}[t]
	\centering
	\subfigure[Landscape Map]{
		\includegraphics[height=2.6cm]{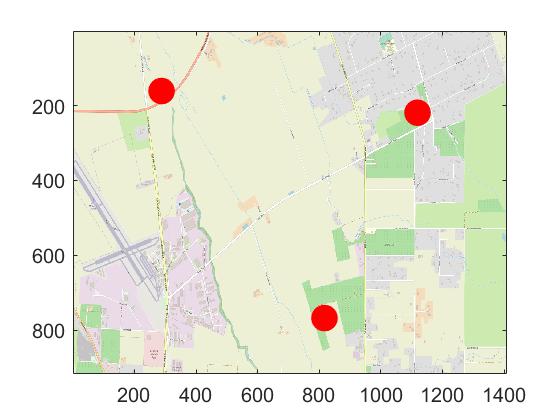}}
  \hspace{3mm}
	\subfigure[Radiomap]{
		\includegraphics[height=2.6cm]{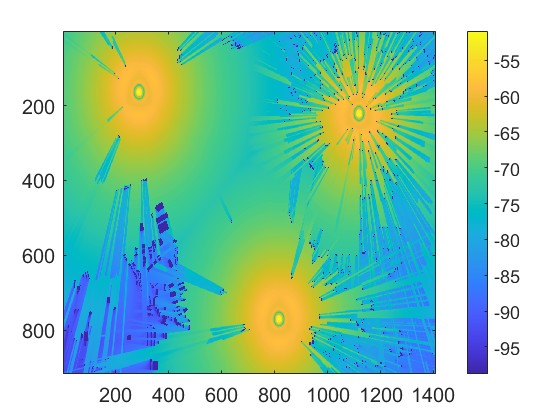}}
	\subfigure[Sparse Observations]{
 \label{sp_ex}
 \hspace{3mm}
		\includegraphics[height=2.6cm]{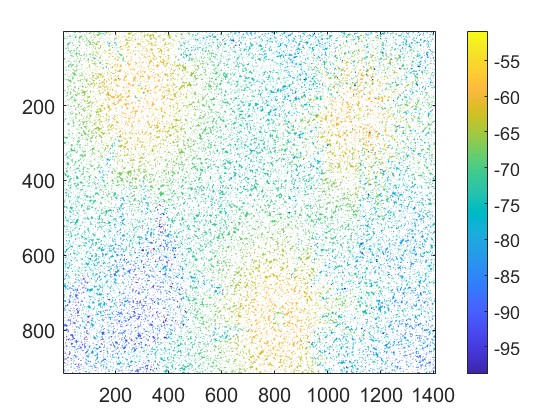}}
    \hspace{3mm}
	\subfigure[Restricted Regions]{
  \label{sp_re}
		\includegraphics[height=2.6cm]{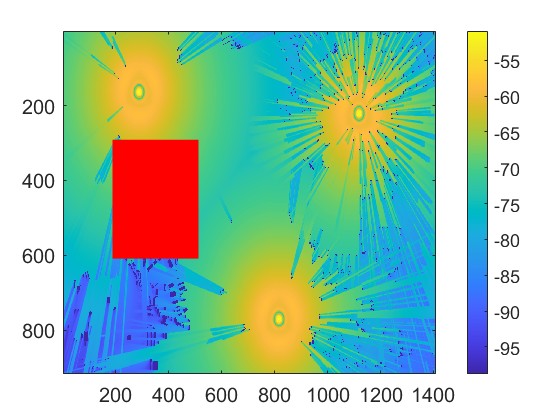}}
	\caption{Examples of radiomap: Fig. a) and Fig. b) show the landscape map and the simulated radiomap ({dBm}) of a selected region with three transmitters, respectively. The transmitter locations are marked as red in Fig. a). The buildings are marked as blue in Fig. b). Fig. c) describes the sparse observations collected from the deployed sensors. Fig. d) shows the radiomap with a restricted region marked as red. Note that, the coordinates here are the index of the signal power strengths conformed to the grid ({$5\times5$ meters$^2$ for each grid block.}).
	}
	\label{rm_Ex}
\end{figure}

\IEEEPARstart{W}{ith} the rapidly expanding deployment
of Internet-of-Things (IoT) sensors and 5G devices, radio spectrum information is becoming more complex,
dynamic, and harder to
measure, thereby posing
vital challenges in spectrum planning and network coverage analysis \cite{c1}. Spectrum resource management and network planning 
critically rely
on accurate information of spatial
radio frequency (RF) signal distribution and coverage.
To this end, radiomaps {play important roles} in modern wireless communication infrastructures. ``Radiomap" 
characterizes  {distribution of power spectral density (PSD), resulting from concurrent wireless signal transmissions, as a function of spatial position, frequency, and time \cite{c2}, where each pixel describes the spectrum power strength measurement shown as Fig. \ref{rm_Ex}.} 
Providing vital and rich information regarding spectrum patterns and RF activities, radiomaps have inspired and assisted
massive applications, including celluar network fault diagnosis \cite{c3}, unmanned aerial vehicle (UAV) path planning \cite{c4}, and autonomous driving \cite{c5}. Practically, high-resolution radiomap is often constructed from observations collected by sparsely deployed
sensors or user devices, shown as Fig. \ref{sp_ex}. Thus, one practical challenge lies in the efficient and accurate reconstruction of more complete radiomaps from partially observed sparse signal power samples.

Recent proposals for radiomap reconstruction can be categorized as either model-based or data-based (i.e.,model-free) approaches. Model-based methods usually assume a specific model for radio propagation between the receivers and transmitters. For example, a log-distance path loss (LDPL) model in \cite{c6} helps estimate the spatial distribution of WiFi radio strengths. In \cite{c7}, another model-based algorithm is based on the thin-plate kernels for radiomap estimation. By contrast, model-free data-based methods estimate the radiomaps relying on observed PSD patterns without assuming a  propagation model. Classic methods include inverse distance weighted (IDW) interpolation \cite{c8} and Radial Basis Function
(RBF) interpolation \cite{c9}. Beyond interpolation-based methods, learning approaches, such as Unet \cite{c10} and generative
adversarial network (GAN) \cite{c11}, have also shown promises in radiomap reconstruction. Other
methods for radiomap estimation include image inpainting \cite{c12}, graph signal processing \cite{c13} and tensor completion \cite{c14}. We provide a more detailed overview in Section \ref{related}.

Despite various successes, most existing approaches to radiomap reconstruction from sparse observations assume sufficient samples of received signal power strengths from sensors 
preferably spread across the whole region as shown in Fig. \ref{sp_ex}. 
However, off-limit or restricted regions, such as mountain ranges and/or private properties, are inaccessible
to radiomap measurements because of physical constraints or security consideration. As illustrated in Fig. \ref{sp_re}, when radio measurements are not available, a radiomap would contain
blanked {areas}. Radio spectrum strength estimation in such restricted areas can be
much more challenging and sometimes impractical for traditional radiomap construction methods. 
First, unlike in coarse radiomaps, missing observations often cover
restricted areas that are important
to fine-resolution radio planning. 
Radio coverage in these smaller areas are
more sensitive to landscape
specifics instead of following a classic radio propagation model, making model-based methods ineffective. On the other hand,  unevenly sampled measurements and blank {areas} on a radiomap can lead to less dependable  
RF power estimate in the {objective regions}, particularly for the locations near the center of the restricted regions, 
which further limits the efficacy of interpolation-based method. 
In addition, various practical reasons
make it harder to collect {large numbers
of measurement data samples},  leading to insufficient training samples for supervised learning algorithms. 
Another analogous problem to radiomap reconstruction with missing {areas} is that of image inpainting \cite{c15}.
However, the lack of regional radio propagation information leads to 
poor performance when imitating
image-inpainting approaches
to recovery accurate radio
spectrum patterns.

This work focuses on capturing spatial spectrum power distribution from limited and uneven observations. In particular, 
we develop a radio depth map and integrate radio propagation model with image inpainting for radiomap reconstruction covering blank areas. 
{We propose two inpainting algorithms 
to address small-size fine-resolution radiomap reconstruction and large-size radiomap reconstruction, respectively.}
The main contributions are:
\begin{itemize}
    \item Exploring radio spectrum 
    patterns and integrating radio propagation models, we propose an exemplar-based method for small-size fine-resolution radiomap estimation with a novel radio-based inpainting priority.
    \item Based on analysis of urban landscape {combined} with the observed radiomaps, we define a novel radio depth map and introduce a two-step template-perturbation method for large-size blank inpainting.
    \item We provide two high-fidelity simulated datasets, where the superior experimental results demonstrate the efficacy of the proposed radiomap inpainting algorithms.
\end{itemize}

We organize the rest of this
manuscript as follows. Section \ref{related}
presents an overview of related works on radiomap reconstruction and image inpainting. Following 
our problem statement in Section \ref{description}, we propose an exemplar-based radiomap inpainting using propagation priority in Section \ref{exemplar}.
We then present our two-step template-perturbation radiomap reconstruction with {a newly-defined radio depth map} in Section \ref{depth}.  Experimental results in Section \ref{results} demonstrate the
performance of the proposed radiomap
{reconstruction} before the conclusions in Section \ref{concolusion}.

\section{Related Works} \label{related}
Here we provide an overview of radiomap {reconstruction/estimation} and image inpainting.
\subsection{Radiomap Reconstruction}
\subsubsection{Model-based Radiomap Estimation}
Model-based approaches usually assume a certain radio propagation model. Specifically, power spectrum can be modeled as a function of frequency $f$ and
position coordinate $\mathbf{c}$ by
\begin{equation}
    r(\mathbf{c},f)=\sum_{i=1}^{N_t}g_i(\mathbf{c},f) s_i(f),
\end{equation}
where $N_t$ is the number of transmitters, $g_i$ denotes the channel power gain of the $i$th transmitter, and $s_i(f)$ is the PSD from the $i$th transmitter \cite{c2}. {Considering different spectrum usage patterns in various frequency bands, one may usually focus on the radiomap reconstruction in a certain frequency $f_j$ from sparse observations \cite{c2}.} A typical example of model-based interpolation is based on log-distance path loss model (LDPL), which has been successefully applied in the scenario of single-narrowband WiFi \cite{c6}. Other model-based methods include kernel expansion \cite{c7}, parallel factor analysis \cite{c16} and fixed rank kriging \cite{c17}.

\subsubsection{Model-free radiomap Estimation}
Unlike model-based radiomap estimation, model-free methods do not suppose a specific radio propagation model but leveraging neighbor spectrum observations. Furthermore, model-based approaches include interpolation  methods and deep-learning (DL)-based methods.

Interpolation methods express the radio spectrum power at a particular location as a combination of observed measurements, denoted by
\begin{equation}
    r(\mathbf{c},f)=\sum_{i=1}^{N_s} w_i(\mathbf{c},f)q_i(f),
\end{equation}
where $q_i$ is the observation from the $i$th receiver and $w_i$ is the combination weights \cite{c2}. For example, in \cite{c8}, the weight $w_i$ can be defined by the inverse distance between transmitters and receivers for the wireless localization. In addition to linear interpolation, an alternative is radio basis function (RBF) interpolation \cite{c18}, where different kernels, such as Gaussian or spline, can be applied. Beyond traditional interpolation approaches, recent works of radiomap reconstruction take advantages of novel data analysis techniques, such as graph signal processing \cite{c13}, multicomponent optimization \cite{c20}, matrix completion \cite{c21} and ordinary
Kriging \cite{c22}.

DL-based approaches has recently attracted significant interest for radiomap estimation, owing to their power in capturing underlying data features and/or mapping functions. Different from interpolation methods, DL-based methods tend to directly find the mapping between the geometric landscape map and the spectrum power measurements through training. Usually, such a functional relationship can be represented by a neural network. For example, Unet is introduced in \cite{c10} for pathloss prediction. Another well-known architecture, namely autoencoder, also
demonstrated its strength in radiomap estimation \cite{c23}. Other learning-based approaches include transfer learning \cite{c24}, GAN-based frameworks \cite{c11,c25,c26,c27}, reinforcement learning \cite{c29} and deep nerual networks \cite{c30}.

Interested readers may refer to the review paper in \cite{c1} for more details. Although radiomap estimation has seen years of exploration, most of existing approaches focus on the reconstruction of radiomap from sparse observations, without considering
the practical issue of missing measurement in restricted areas.

\subsection{Image Inpainting}
\begin{figure}[t]
    \centering	
     \includegraphics[height=3cm]{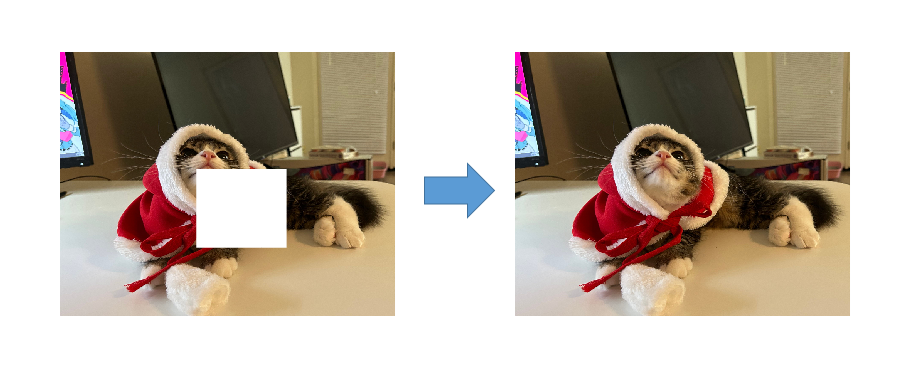}
     \caption{Example of Image Inpainting.}
     \label{example:cat}
\end{figure}

Image inpainting is the process 
for completing missing regions in images or for removing foreign objects added to more natural images \cite{c31}, as shown in Fig. \ref{example:cat}. A classic method is exemplar-based image inpainting, which
reconstruct missing regions 
from selected exemplar patches \cite{c32}. Other modern concepts, such as dictionary learning \cite{c33}, subspace analysis \cite{c34} and information diffusion \cite{c35}, can also improve exemplar-based inpainting. Recently, deep learning has 
demonstrated many successes in computer vision tasks, as well as in image inpainting. Typical learning frameworks include deep convolutional neural networks \cite{c36,c37,c38}, auto-encoder \cite{c39} and generative adversarial networks \cite{c40,c41,c42}.

Despite the many successes of exemplar-based and learning-based inpainting approaches in generic image processing, existing methods have only shown modest success in radiomap reconstruction, possibly because they
tend to neglect the fundamentals of
radio physics. Moreover, since the radiomaps with missing patches often correspond to certain locations with specific local landscape features, there may not 
exist enough  samples for effective
training of DL based on neural networks.
How to efficiently inpaint the radiomap still remains an open challenge.

\section{Problem Description} \label{description}
\begin{figure}[t]
    \centering	
     \includegraphics[height=2.8cm]{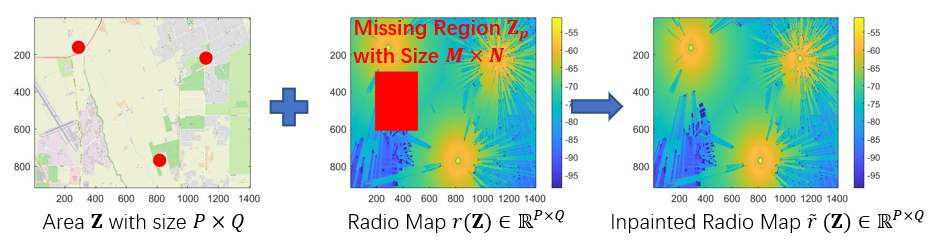}
     \caption{Illustration of Objective Scenarios: A restricted area $\mathbf{Z}_p$ with size $M\times N$ is located in a larger area $\mathbf{Z}$ with size $P\times Q$. Given the observations of unrestricted areas in the radiomap $r(\mathbf{Z})$ together with the landscape, our goal is to estimate the signal power $\tilde{r}(\mathbf{Z}_p)\in\mathbb{R}^{M\times N}$ in the restricted areas to complete the radiomap $\tilde{r}(\mathbf{Z})\in\mathbb{R}^{P\times Q}$ of the whole region.}
     \label{example:problem}
\end{figure}

\begin{figure}[t]
    \centering	
     \includegraphics[height=2.8cm]{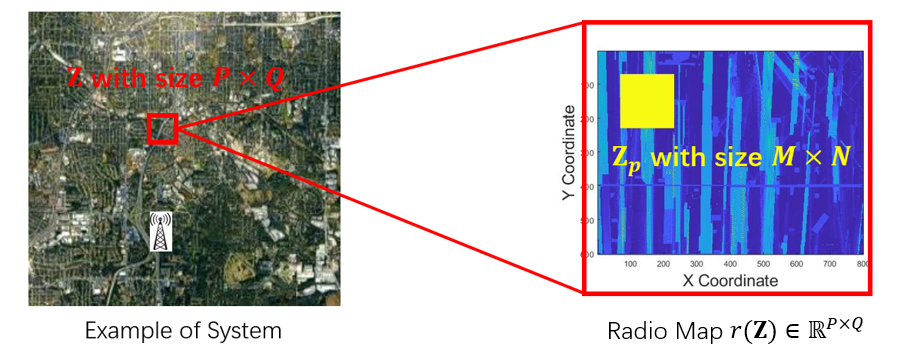}
     \caption{Example of fine-resolution radio in a small-scale region (JHU-APL Dataset \cite{c12}):  Restricted area is marked yellow.}
     \label{example:small}
\end{figure}

We study a wireless network coverage of a rectangular area corresponding to an
available landscape map. Without loss of
generality, our {radio spectrum power observations} are located in a rectangular area $\mathbf{Z}$ with size $P\times Q$, and are arranged in a regular grid, shown as Fig. \ref{example:problem}. We denote the radiomap of $\mathbf{Z}$ by $r(\mathbf{Z})\in\mathbb{R}^{P\times Q}$, where each  power observation $r(Z_i)$ (typically measured in dBm) is characterized by a 2-dimensional coordinates $Z_i=(x_i,y_i)$. A restricted area $\mathbf{Z}_p$ with size $M\times N$ is located within $\mathbf{Z}$, marked as red in Fig. \ref{example:problem}, where $M\leq P$ and $N\leq  Q$. No measurement in $\mathbf{Z}_p$ is available. Our objective is to estimate the radiomap $\Tilde{r}(\mathbf{Z}_p)$ of the missing regions from other observed samples, with the consideration of local landscape, such as buildings and roads, and locations of transmitters. More specifically, we explore different radiomap inpainting methods for two scenarios:

\begin{itemize}
    \item Small-scale fine-resolution radiomap: Compared to traditional radiomap reconstruction problems, the small-scale radiomap has higher resolution (e.g., down to 1 meter) and smaller area, which make it more sensitive to the nearby landscape, such as buildings, trees and roads, as shown in Fig. \ref{example:small}. Moreover, these regions usually have limited and unbalanced observations around restricted/inaccessible regions. We  address the reconstruction of such small-scale fine-resolution radiomaps in Section \ref{exemplar}.
    \item Large-scale radiomap: Unlike small-scale radiomap inpainting,  large-scale radiomap estimation favors radio propagation patterns with the consideration of shadowing and landscape, shown as Fig. \ref{example:problem}. To capture both landscape and model information efficiently is a challenge. Large-scale radiomap inpainting is to be discussed in Section \ref{depth}.
\end{itemize}

\section{Exemplar-based radiomap Inpainting with Propagation Priority} \label{exemplar}

We now introduce an exemplar-based radiomap inpainting using radio propagation priority for small-scale {radiomaps}, where neighboring landscape dominates the radio spectrum patterns.

\begin{figure*}[t]
	\centering
	\includegraphics[width=6in]{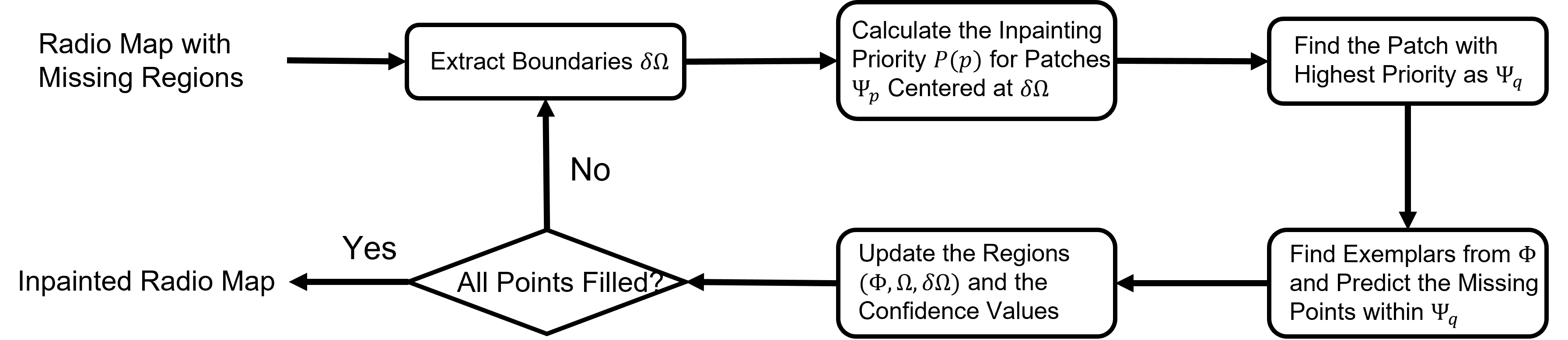}
	\caption{Scheme of Proposed Method}
	\label{sch}
\end{figure*}

\begin{figure}[t]
	\centering
	\subfigure[]{
		\label{exem1}
		\includegraphics[height=2.5cm]{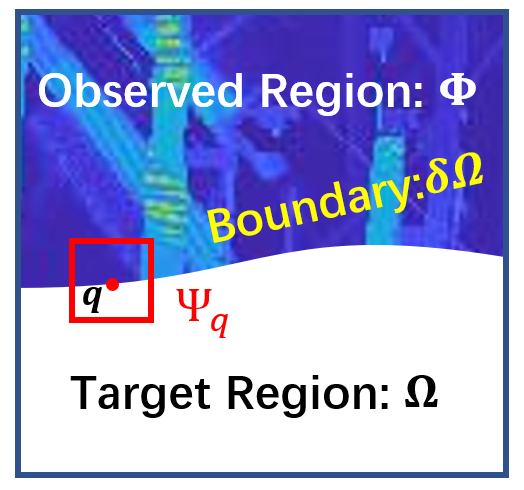}}
	\hspace{2cm}
	\subfigure[]{
		\label{exem2}
		\includegraphics[height=2.5cm]{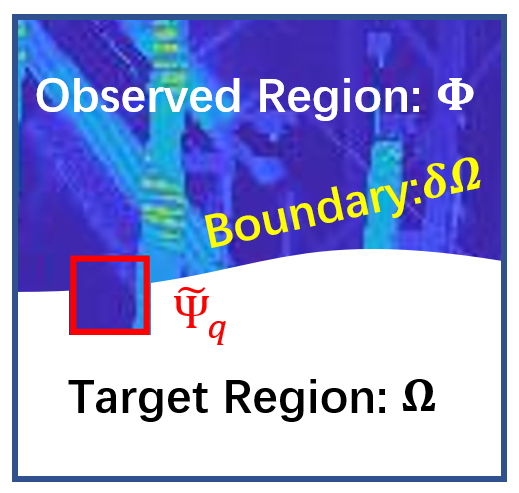}}
	\caption{Illustration of Filling Process: a) Select a patch $\Psi_q$ in the boundary $\delta \Omega$; b) Estimate the missing values in $\Psi_q$ and regenerate $\tilde\Psi_q$}
	\label{exeme}
\end{figure}

\subsection{Overview of the Proposed Method}
To fill a region based on 
surrounding observations, one intuitive way is to estimate missing values patch-by-patch (or block-by-block) from boundaries between observed regions and target (restricted area) regions, leading to the center of the restricted/inaccessible area, by following a scheme named as exemplar-based inpainting \cite{c32} illustrated in Fig. \ref{sch}. 

Refining notation for the observed regions in $\mathbf{Z}$ as $\mathbf{\Phi}$ and the restricted region $\mathbf{Z}_p$ by $\mathbf{\Omega}$, we are able to detect the boundary between 
$\mathbf{\Phi}$ and $\mathbf{\Omega}$ as $\delta\Omega$. In the exemplar-based inpainting, a patch (block) of interest centered at the boundary $\delta \Omega$ could be selected for estimation first, as shown in Fig. \ref{exeme}. Suppose that $\Psi_p$ is a $n\times n$ patch centered in a location $p$ located at boundaries, i.e., $p\in\delta \Omega$. One can calculate  inpainting priority of the patch $\Psi_p$ as $P(p)$ based on texture properties and neighbor measurement information. Ordering all patches centered at $\delta \Omega$ by $P(p)$, a patch $\Psi_q$ with the highest priority and the most significant interest can be selected. Next, several exemplar patches can be extracted from the observed regions $\mathbf{\Phi}$ depending the similarity to $\Psi_q$, after which the missing measurement in $\Psi_q$ is reconstructed from the exemplar patches. Once the selected $\Psi_q$ is filled, we could update the regions $\{\mathbf{\Phi}, \mathbf{\Omega}, \delta \Omega\}$ and repeat the aforementioned steps until missing
information in the restricted area $\mathbf{Z}_p$ are estimated.

From the steps above, key issues in the exemplar-based method are to define priority $P(p)$ to determine the filling direction, and to estimate the missing values from exemplars. In the following subsections, we will discuss our proposed methods on the integration of radio propagation model and exemplar-based inpainting for radiomap estimation.

\subsection{Details of Proposed Method}
In this part, we introduce a novel inpainting priority based on radio propagation model and two approaches to estimate missing radiomap values.

\subsubsection{Definition of Priority}
{We start with the scenario with one single transmitter.}
To determine the proper direction of inpainting the missing values, we expect to propagate the important patterns in texture ({distribution of spectrum power}) and radio strength measurements with larger certainty. Thus, we could define the following priority of patch selection:
\begin{figure}[t]
	\centering
	\subfigure[Data scalar.]{
		\label{pp1}
		\includegraphics[height=3.5cm]{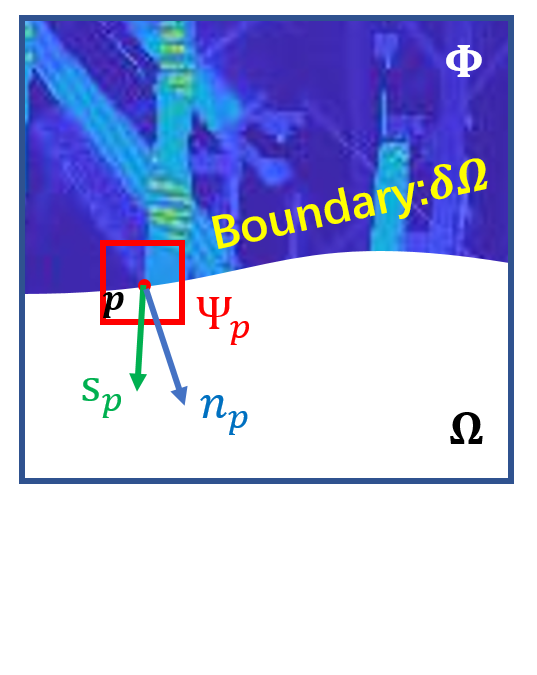}}
  \hfill
	\subfigure[Radio factor.]{
		\label{pp2}
		\includegraphics[height=3.5cm]{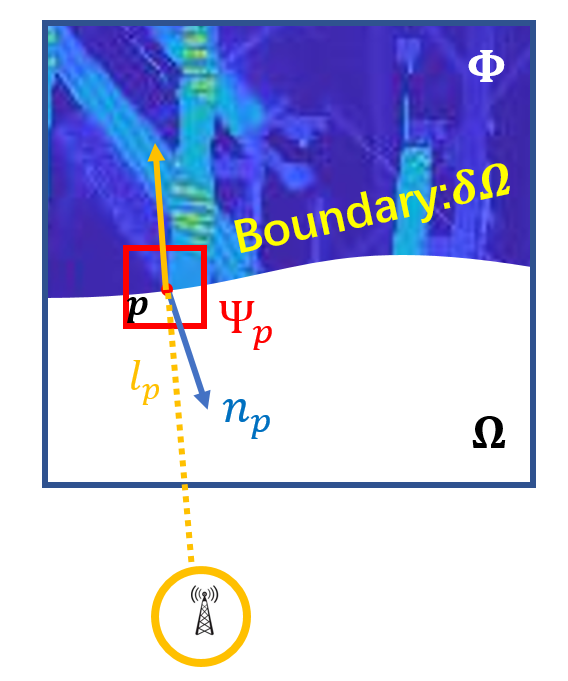}}
	\hfill
	\subfigure[Block term.]{
		\label{pp3}
		\includegraphics[height=3.5cm]{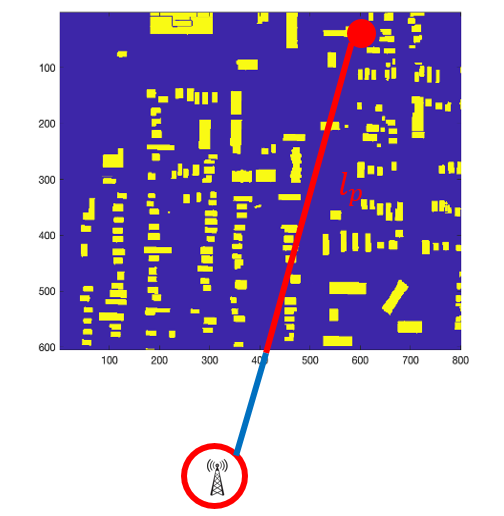}}
	\caption{Illustration of Calculating Priority}
	\label{pp}
\end{figure}
\begin{equation}\label{priority}
	P(p)=C(p)\cdot D(p)\cdot B(p)\cdot L(p),
\end{equation}
where confidence $C(p)$ and data scalar $D(p)$ contain radiomap pattern information (texture), while radio propagation factor $L(p)$ and block term $B(p)$ describe radio propagation properties. More specifically, each term can be defined as follows:
\begin{itemize}
\item $C(p)$: Confidence $C(p)$ represents how confident the values in $\Psi_p$ are. If more samples are observed from $\mathbf{\Phi}$, the corresponding patch is with a larger confidence term. {Suppose that the patch size is $n\times n$}. The confidence term can be written as 
\begin{equation}\label{pri}
    C(p)=n^{-2} {\sum_{v \in (\Psi_p\cap \Phi)}C(v)},
\end{equation}
where $C(v)$ is initialized as $C(v)=1$ for $v\in \Phi$; otherwise, $C(v)=0$. {Here, $C(p)\in[0,1]$ describes the averaged confidence levels in the $\Psi_p$}. Note that, at each round, the confidence term $C(u)$ is updated by $C(u)=C(q)$ for a newly-estimated position $u$ before the next round.
\item $D(p)$: Data scalar $D(p)$ focuses on texture patterns near patch $\Psi_p$, shown in Fig. \ref{pp1}. More specifically, it should describe diffusion of textures and favor local gradients. Let $\mathbf{n}_p$ be the normal of boundary at $p$. Then 
we can denote the orthogonal direction of the texture gradient at $p$ as $\mathbf{s}_p=[\nabla T_p]^{\perp}$ where $T_p$ is the power level around $p$. We define data scalar
	\begin{equation}\label{dat1}
		D(p)=\frac{|<\mathbf{s}_p, \mathbf{n}_p>|}{\|\mathbf{s}_p\| \|\mathbf{n}_p \| |},
	\end{equation}
where $<\cdot, \cdot>$ denotes inner product. Such data scalar describes the {local variations/gradients of radiomaps in the direction hitting the boundaries.}
 \item $L(p)$: Radio propagation factor captures the radio model pattern between the received signal strenght at location $p$ and the transmitter at location $p_t$. Similar to the data term $D(p)$, we prefer to diffusing the radiomap patterns following the direction of radio propagation. Thus, we could measure the strengths of radio hitting the boundary $\delta \Omega$ by
 \begin{equation}\label{eq:radio_model}
     	L(p)=|d(p_t,p)|^{-\beta}|\mathbf{l}_p\cdot \mathbf{n}_p|,
 \end{equation}
 where $|d(p_t,p)|$ is the distance between transmitter position $p_t$ and receiver location $p$, $\beta$ is a hyperparameter capturing the inverse distance weights  (IDW) \cite{c8}, $\mathbf{n}_p$ is the normal of boundary at $p$, and $\mathbf{l}_p$ is the direction of radio propagation from $p_t$ to $p$, shown in Fig. \ref{pp2}. Note that, here we use IDW model as an example. Other options, such as LDPL, could also be applied.
 \item $B(p)$: In small-scale areas, radiomap around an off-limit area is usually sensitive to landscape differences. Thus,
we define a block term $B(p)$ to capture the shadowing and fading from obstacles.
 From the urban landscape map, we can segment buildings (in yellow) and background/{free-space} (in blue) as shown in Fig. \ref{pp3}. Suppose that $l_p$ be 
the segment of the path linking $t$ and $p$, i.e., red areas in Fig. \ref{pp3}. Then, $B(p)$ is defined as 
	\begin{equation}\label{btm}
		B(p)={\textnormal{fraction of cumulative non-building length in } l_p}
	\end{equation}
If  radio propagates over more obstacles, $B_p$ is smaller and the priority would be lower.
\end{itemize}
Note that, thus far, our examples focus mainly on one transmitter cases. 
Under $N_t$ simultaneous transmitters, the radio-based priority factor in Eq. (\ref{priority}) 
can be modified as
\begin{equation}\label{summ}
	P(p)=C(p)\cdot D(p)\cdot [\sum_{i=1}^{N_t} B_i(p)\cdot L_i(p)],
\end{equation}
where $B_i(p)$ and $L_i(p)$ are respectively
defined for the $i$th transmitter. 
{Note that, we consider all transmitters of identical
settings. One may also apply weighted sum in Eq. (\ref{summ}) according to the relative strengths of each transmitter, which can be embedded in the term $L_i(p)$ via LDPL models.} By choosing 
patches with larger $P(p)$, we would
optimize the inpainting direction by considering both textures and radio propagation.

\subsubsection{Recovery of Missing Measurement}
After selecting patch $\Psi_q$ with the highest priority, we now discuss the problem of missing measurement recovery in off-limit regions. Here, we introduce two exemplar-based approaches, i.e., EPC and EPD, as follows:

\begin{itemize}
	\item Estimation via Exemplar copy (EPC): A typical appraoch to estimate the missing points is to copy values from similar observed patches of the same indices \cite{c15}. Here, we also consider exemplar-based copying to estimate the missing points. Let $\Psi_q$ be the $n\times n$ patch selected by $P(p)$. {Here, $n$ can be the hyperparameter selected based on the specific region size and data features}. To implement EPC, we first find the closest exemplar patch $\Psi_s$ from the observed region according to
	\begin{equation}
		\Psi_s=\arg \min_{\Psi_w, w\in \Phi}\sum_{i\in \Phi}[(\Psi_w)_i-(\Psi_q)_i]^2,
	\end{equation}
	where $(\Psi)_i$ is the signal strength measurement at position $i$ within patch $\Psi$. We then fill the missing value via
	\begin{equation}\label{epc}
		(\tilde \Psi_q)_i=\left \{
		\begin{aligned}
		(\Psi_q)_i	&\quad \quad i\in\Phi\\
			(\Psi_s)_i&\quad\quad i\in\Omega
		\end{aligned}
		\right..
	\end{equation}
	\item Estimation via dictionary learning (EPD):
        Dictionary learning 
        in exemplar-based image inpainting
        \cite{c33} optimizes a sparse vector to combine codewords in a dictionary.  For an $n\times n$ patch $\Psi_q$, we may randomly pick a set of patches from $\Phi$ and generate a dictionary $\mathbf{A}\in\mathbb{R}^{n^2 \times K}$ containing $K$ normalized code-words via K-SVD \cite{c43} or matching pursuit \cite{c44}. Reshaping patch $\Psi_q$ as a vector $\mathbf{x}_q$, dictionary learning optimizes the sparse vector $\bm{\beta}\in\mathbb{R}^{K\times 1}$ for codeword combining via:
	\begin{equation} 
		\tilde{\bm{\beta}}=\arg \min_{\bm{\beta}} ||(\mathbf{x}_q)_\Phi-\mathbf{A}_\Phi \bm{\beta}||^2_2+\lambda||\bm{\beta}||_1,
	\end{equation}
	where
 $(\mathbf{x}_q)_\Phi$ is an observed part in $\Psi_q$ and {$\mathbf{A}_\Phi \bm{\beta}$ are the encoded signals corresponding to $(\mathbf{x}_q)_\Phi$}. The optimized $\bm{\beta}$ allows
 radiomap reconstruction for missing regions:
	\begin{equation} \label{epd}
	(\tilde \Psi_q)_i=\left \{
	\begin{aligned}
	(\Psi_q)_i	&\quad \quad i\in\Phi\\
	(\mathbf{A}\bm{\beta})_i&\quad\quad i\in\Omega
	\end{aligned}
	\right..
	\end{equation}
\end{itemize}
In general, EPC performs better when radiomap contains regular and smooth patterns, whereas EPD 
is more adept in handling complex landscape. Exemplar-based radiomap inpainting with propagation priority can be summarized in Algorithm \ref{alg:A}.

\begin{algorithm}[t]
    \caption{Exemplar-based Radiomap Inpainting with Radio Propagation Priority (EPC/EPD)}
    \label{alg:A}
    \begin{algorithmic}
        \STATE {\textbf{Input}}: Radiomap $r(\mathbf{Z})\in\mathbb{R}^{P\times Q}$ and the restricted area $\mathbf{Z}_p$ with size $M\times N$ (observed region is denoted by $\Phi$ while missing region $\Omega$ is initialized by $\mathbf{Z}_p$), and building segmentation $m(\mathbf{Z})\in\mathbb{R}^{P\times Q}$.\\
        \textbf{while} $\Omega\neq \emptyset$ \textbf{do}:
        \STATE \quad{\textbf{1.} Extract the boundary $\delta \Omega$ between observed region $\Phi$ and target region $\Omega$;}
        \STATE{\quad\textbf{2.}} Given a patch $\Psi_p$ with size $n\times n$ centered at point $p$ located at boundaries, i.e., $p\in\delta \Omega$, calculate the priority of the patch as $P(p)$ based Eq. (\ref{priority});
        \STATE{\quad\textbf{3.}} Order all patches $\Psi_p$ centered at $\delta \Omega$ by $P(p)$ and select the one with highest priority as $\Psi_q$;
        \STATE{\quad\textbf{4.}}  Select exemplars from observed regions for $\Psi_q$ and estimate the missing values in $\Psi_q$ by EPC in Eq. (\ref{epc}) or EPD in Eq. (\ref{epd});
        \STATE{\quad\textbf{5.}} Update $\Phi$ and $\Omega$;
        \STATE{\quad\textbf{6.}} Update the confidence $C(p)$ term in the priority as calculated in Eq. (\ref{pri});\\
        \textbf{end while}
        \STATE{\textbf{Output}:} Estimated radiomap $\Tilde{r}(\mathbf{Z}_p)$ of the protected area.
    \end{algorithmic}
\end{algorithm}

\section{Template-perturbation radiomap Inpainting based on Radio Depth Map} \label{depth}

In this section, we introduce a two-step template-perturbation radiomap inpainting for more general large-scale regions, as shown in Fig. \ref{lexeme}.

\subsection{Large-scale radiomap Inpainting}
\begin{figure}[t]
	\centering
	\subfigure[]{
		\label{lexem1}
		\includegraphics[height=2.8cm]{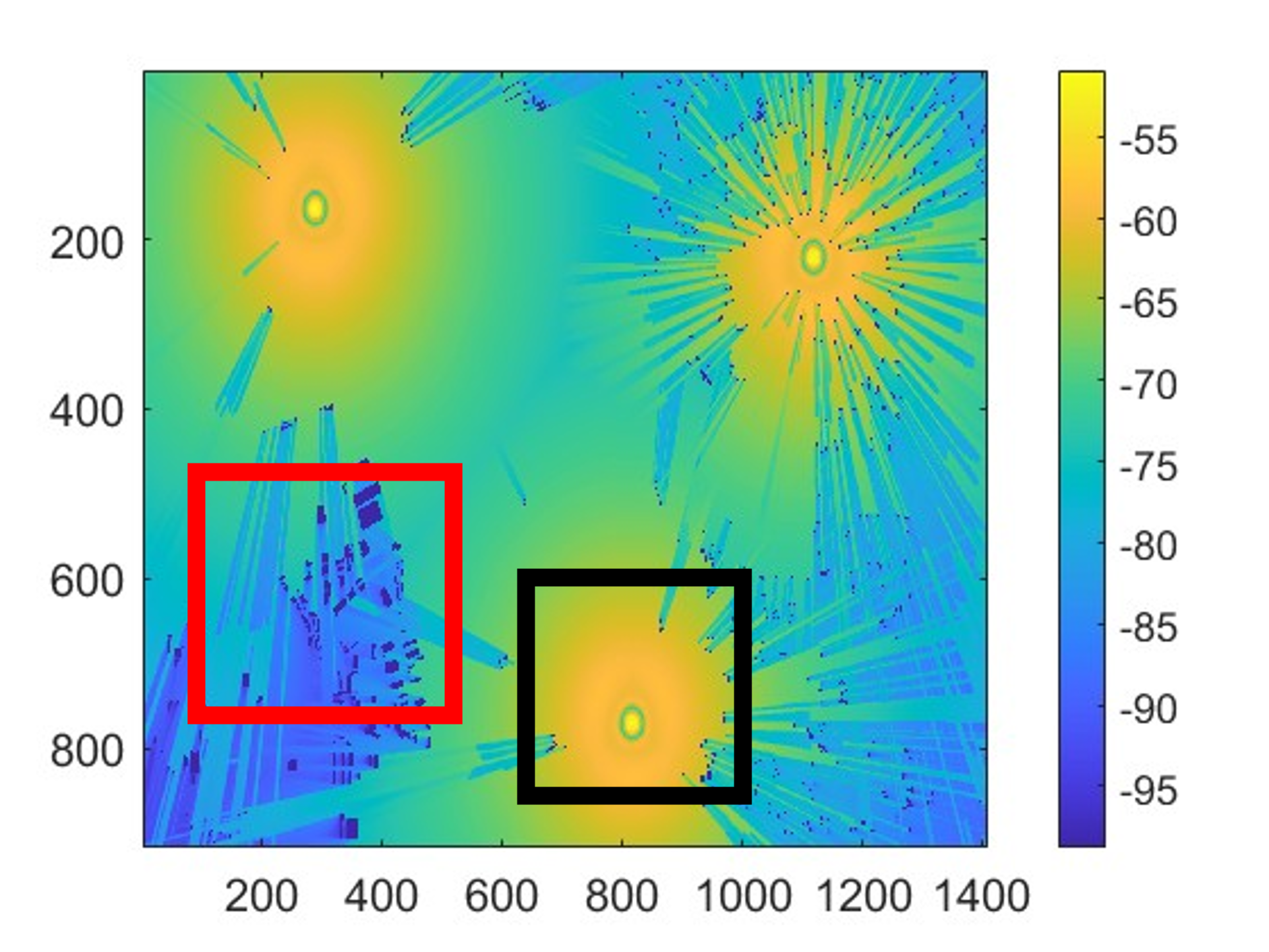}}
	\hspace{0.5cm}
	\subfigure[]{
		\label{lexem2}
		\includegraphics[height=2.8cm]{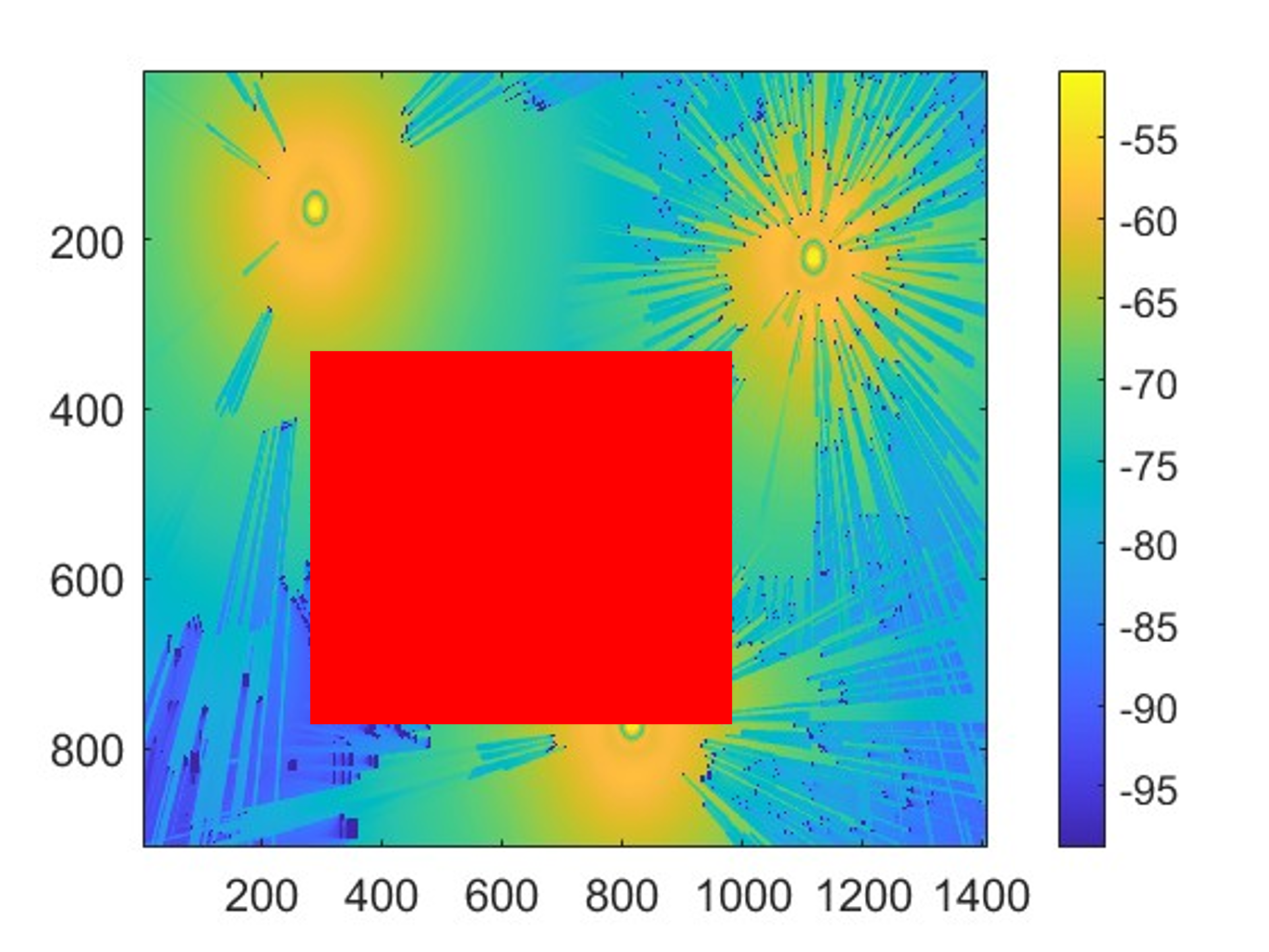}}
	\caption{Example of large-scale regions of radiomap (dB): a) Illustration of spectrum patterns in different regions, where red parts favors the nearby landscape and black parts is dominated by the propagation model; b) Example of large-size protected areas, which lack the spectral neighbor information near the center of the region.}
	\label{lexeme}
\end{figure}

Large-sacle radiomap inpainting faces additional challenges beyond small scale radiomap:

\begin{itemize}
    \item Small-scale radiomap is sensitive to surrounding landscape, whereas certain regions in
    large-scale radiomap may be better characterized by large-scale propagation model. For example, as shown in Fig. \ref{lexem1},  spectrum patterns in the black block 
    are
    dominated by large scale propagation model whereas 
    PSD distribution within the red block is more sensitive to nearby landscape. 
An important problem is to effectively capture both radio propagation and local shadowing effect. {In our proposed EPC/EPD, we embed these landscape information in the propagation priority defined by Eq. (\ref{btm}).}
    \item In many cases involving low {sensor population density},
    missing measurement regions on a radiomap may cover a larger area and
    do not have many observations for radiomap reconstruction especially for the middle
    of large regions,  as shown in Fig. \ref{lexem2}. For radiomap recovery of
    larger-scale restricted regions,
computational complexity and estimation accuracy also present additional challenges.  
\end{itemize}

\begin{figure*}[t]
	\centering
	\includegraphics[width=6in]{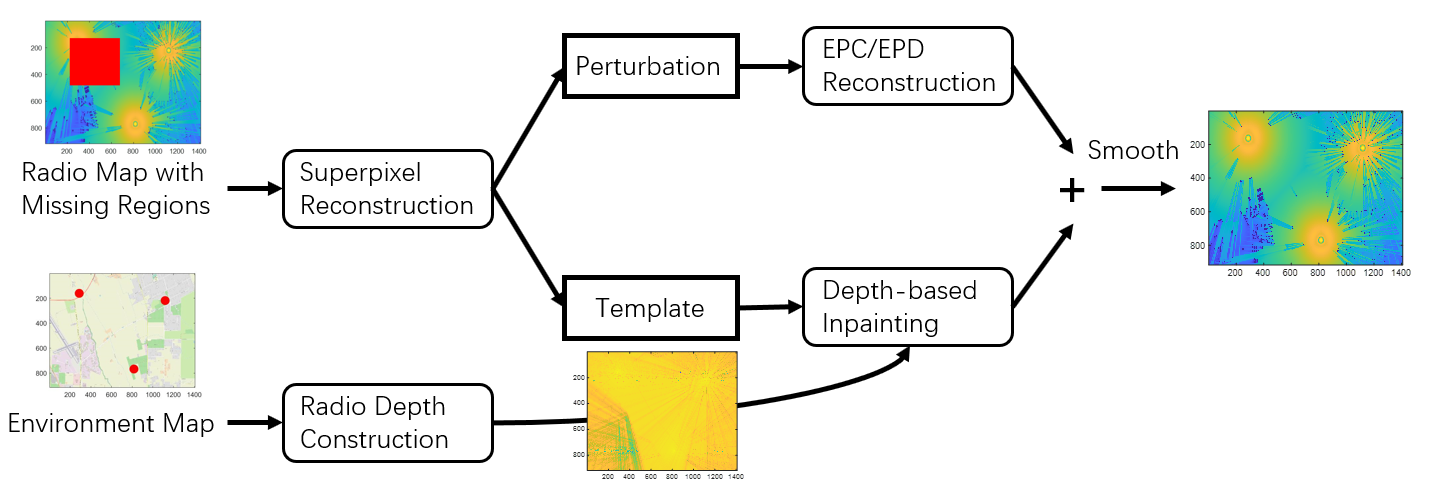}
	\caption{Scheme of Proposed Two-Step radiomap Inpainting Method}
	\label{sch2}
\end{figure*}

\begin{figure}[t]
    \centering	
     \includegraphics[height=3cm]{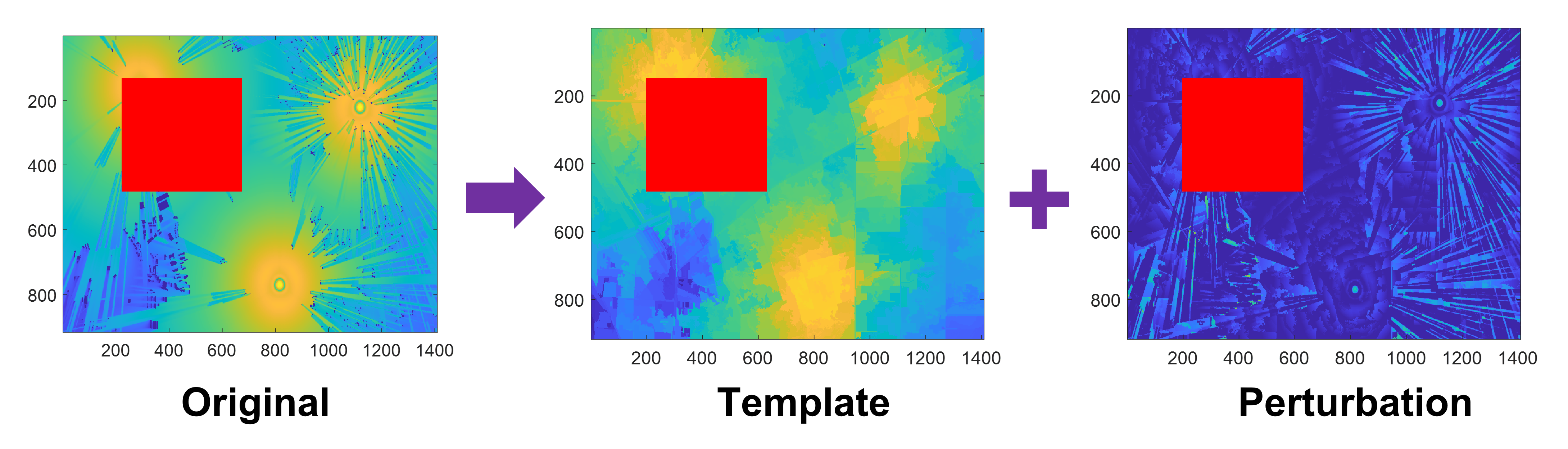}
     \caption{Example of radiomap Decomposition.}
     \label{example:decom}
\end{figure}

To address the aforementioned challenges, here we propose a two step template-perturbation radiomap inpainting. More specifically, we decompose the original radiomap into a low-resolution template 
capturing the overall radio propagation pattern, and a high-resolution perturbation characterizing detailed
shadowing effects as shown in Fig. \ref{example:decom}. For the high-resolution perturbation, we apply the exemplar-based radiomap inpainting, i.e., EPC/EPD as described in Algorithm \ref{alg:A}. For the low-resolution template, we define a novel radio depth map to assist
propagation template inpainting. The overall strategy of the propose two-step algorithm is 
illustrated in Fig. \ref{sch2}, with additional details discussed next.

\subsection{Details of the Proposed Methods}\label{sec:detail2}

\subsubsection{Construction of Template and Perturbation}
Given a radiomap $r(\mathbf{Z})\in\mathbb{R}^{P\times Q}$ containing restricted area $\mathbf{Z}_p$ of size $M\times N$, we first decompose it into a low-resolution template to capture the smooth radio propagation patterns, together with a fine-resolution perturbation to describe the shadowing details as Fig. \ref{example:decom}. 

To construct a low-resolution smooth template, we first construct superpixels based on the landscape map. In this work, we consider the entropy rate superpixel segmentation (ERS) as \cite{c45}, which is with low complexity and competitive efficiency. In ERS \cite{c46}, a graph $\mathcal{G}=\{\mathcal{V}, \mathcal{E}\}$ is constructed to model the dataset, where the pixels serve as the nodes $\mathcal{V}$ and their pairwise similarities are represented by edges $\mathcal{E}$. Next, we formulate the superpixel segmentation problem
as a subgraph construction:
\begin{align}\label{obj}
&\mathcal{L}^*=\arg_{\mathcal{L}} \max{Tr\{H(\mathcal{L})+\alpha T(\mathcal{L})\}}\\
&s.t. \quad \mathcal{L}\subseteq\mathcal{E},
\end{align}
where $\mathcal{L}\subseteq\mathcal{E}$ is a subset of edges that partition the original graph, entropy rate term $H(\mathcal{L})$ favors more compact clusters, and regularizing term $T(\mathcal{L})$ punishes large cluster size. Such optimization problems can solved by a greedy algorithm. Interested readers could refer to \cite{c46} for more details.

With the superpixel construction, the original region $\mathbf{Z}$ with size $P\times Q$ can be represented by a set of $K$ superpixels, i.e., $\mathcal{S}=\{S_1,S_2,...,S_K\}$ with $\sum_{i=1}^{K}|S_i|=P\cdot Q$. By calculating the mean PSD measurements within the same superpixel as its new value, we are able to construct a low-resolution smooth template of the radiomap. Denote the template
by ${t}(\mathbf{Z})\in\mathbb{R}^{P\times Q}$. Each value in the template located at position $Z_\alpha$ is calculated by
\begin{equation}\label{template_con}
    t(Z_\alpha)=\frac{1}{|S_i|}\sum_{\beta\in S_i}r(Z_\beta),
\end{equation}
where $Z_\alpha\in S_i$. Through calculating the difference between the original radiomap and template, we could obtain the perturbation $h(\mathbf{Z})\in\mathbb{R}^{P\times Q}$ as 
\begin{equation}\label{pert_con}
    h(\mathbf{Z})=r(\mathbf{Z})-t(\mathbf{Z}).
\end{equation}

\subsubsection{Radiomap Inpainting for Template and Perturbation}
Given the definition of perturbation and template, we reconstruct the restricted regions of PSD measurements in $t(\mathbf{Z})$ and $h(\mathbf{Z})$, respectively.

\noindent
\underline{2-A. Perturbation Reconstruction}: Since the perturbation captures detailed shadowing
effects and has similar patterns as fine-resolution small-scale radiomaps, we can easily adopt
the exemplar-based EPC/EPD introduced in Algorithm \ref{alg:A} to reconstruct the missing areas in $h(\mathbf{Z})$. 

\noindent
\underline{2-B. Template Reconstruction}:
We now focus on the exemplar-based inpainting for radiomap template $t(\mathbf{Z})$.
To reconstruct missing values in the template $t(\mathbf{Z})$, we develop an exemplar-based inpainting similar as Fig. \ref{sch} in Algorithm \ref{alg:A}. We reconstruct missing values from the boundary $\delta \Omega$ to the center patch by patch by first defining a $n\times n$ patch $\Psi_p$ centered at location $p\in\delta \Omega$. Next, we calculate all patch priorities $P(p)$ and select the one patch with highest priority as $\Psi_q$ to fill. Subsequently, we select several exemplar patches from the observed areas 
according to their distance to $\Psi_q$, after which we estimate the missing values in $\Psi_q$ from these exemplars. Note that, unlike perturbation inpainting in Algorithm \ref{alg:A} which stresses shadowing
effects and pattern textures, the template spectral patterns rely more on radio propagation model. 

To summarize, we define a novel depth map to calculate the patch priority and a different similarity distance to find the exemplar patches as follows:

\begin{itemize}
    \item \textbf{Radio Depth Map}: In computer vision, a depth map is an image or image channel containing  information related to the distance of surfaces from a viewpoint, often applied for image structure reconstruction \cite{c47}. Motivated by image depth map, we define a novel radio depth map $W(\mathbf{Z})\in\mathbb{R}^{P\times Q}$ as a map containing 
    radio propagation information related to the distance from the viewpoint of transmitters. More specifically, we consider spectrum power
    loss due to the blocking effect from buildings. 
    Let $N_t$ be the number of transmitters. We then calculate
 the depth map value at location $Z_\alpha$ via
    \begin{align}\label{dep}
        W(Z_\alpha)=f(\sum_{i=1}^{N_t} E_i(Z_\alpha)\cdot B_i(Z_\alpha)),
    \end{align}
    where $B_i(\cdot)$ follows the definition of Eq. (\ref{btm}), {$E_i(\cdot)$ captures  power of different transmitters} 
    and $f(\cdot)$ is the normalization function. Note that the propagation factor $E_i(\cdot)$ can be defined by inverse distance weight (IDW) \cite{c8}. Let $d_i$ be the distance between the transmitter $i$ and $Z_\alpha$. Then for a
 hyperparameter $\sigma$, we have
    \begin{equation} \label{dep_idw}
        E_i(Z_\alpha)=d_{i}^{-\sigma}.
    \end{equation}
Another alternative is to use the LDPL model \cite{c6}
    \begin{equation}
    E_i(Z_\alpha)=\theta_i-\epsilon_i\cdot\log(d_i),
    \end{equation}
    where the parameters can be solved by the multi-variable linear regression from the spectrum observations. An example of the depth map in a region with three transmitters are shown as Fig. \ref{EXEMET}. To provide the flexiblility of model tuning and reduce complexity, we usually apply IDW-based depth map for radiomap inpainting.
    \begin{figure}[t]
	\centering
         \subfigure[Transmitter Location]{
		\label{lexem1E}
		\includegraphics[height=2.8cm]{3brat.jpg}}
  \hspace{0.5cm}
	\subfigure[IDW-based]{
		\label{lexem2E}
		\includegraphics[height=2.8cm]{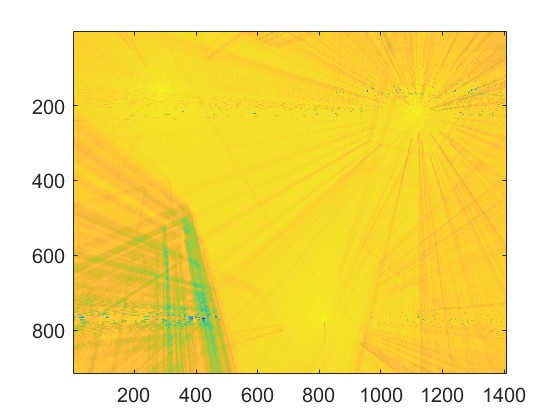}}
	\hspace{0.5cm}
	\subfigure[LDPL-based]{
		\label{lexem3E}
		\includegraphics[height=2.8cm]{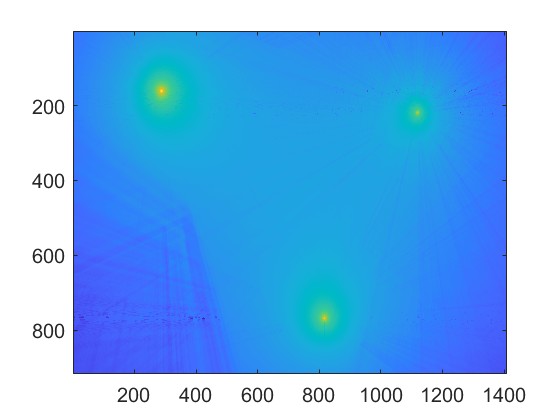}}
	\caption{Example pf Depth Map with Three Transmitters: transmitter locations are marked as red.}
	\label{EXEMET}
\end{figure}
    \item \textbf{Patch Priority}: With a depth map, we now introduce a depth-based priority for radiomap inpainting. Given a patch $\Psi_p$ centered at $p$, we 
define its inpainting priority as
    \begin{equation} \label{pri2}
        P(p)=C(p)\cdot D(p) \cdot V(p),
    \end{equation}
    where confidence term $C(p)$ is defined in Eq. (\ref{pri}), data texture term $D(p)$ is calculated by Eq. (\ref{dat1}), and the depth factor $V(p)$ follows
    \begin{equation}
        V(p)=\frac{|W(\Psi_p)|}{|W(\Psi_p)|+\sum_{q\in\Psi_p\cap \Phi}(W(q)-\overline{W(\Psi_p)})^2},
    \end{equation}
    where $|W(\Psi_p)|$ is the number of observed pixels with patch $\Psi_p$, and its average is denoted by $\overline{W(\Psi_p)}$. $V(p)$ favors patches with smooth patterns and higher depth certainty.
    \begin{algorithm}[t]
    \caption{Exemplar-based Radiomap Inpainting for Template (EPT)}
    \label{alg:B}
    \begin{algorithmic}
        \STATE {\textbf{Input}}: Template radiomap $t(\mathbf{Z})\in\mathbb{R}^{P\times Q}$ and the restricted area $\mathbf{Z}_p$ with size $M\times N$ (observed region is denoted by $\Phi$ while missing region $\Omega$ is initialized by $\mathbf{Z}_p$), and landscape map $m(\mathbf{Z})\in\mathbb{R}^{P\times Q}$.
        \STATE \textbf{1.} Construct the radio depth map $W(\mathbf{Z})\in\mathbb{R}^{P\times Q}$ based on Eq. (\ref{dep}); \\
        \textbf{while} $\Omega\neq \emptyset$ \textbf{do}:
        \STATE \quad{\textbf{2.} Extract the boundary $\delta \Omega$ between observed region $\Phi$ and target region $\Omega$;}
        \STATE{\quad\textbf{3.}} Given a patch $\Psi_p$ with size $n\times n$ centered at point $p$ located at boundaries, i.e., $p\in\delta \Omega$, calculate the priority of the patch as $P(p)$ based Eq. (\ref{pri2});
        \STATE{\quad\textbf{4.}} Order all patches $\Psi_p$ centered at $\delta \Omega$ by $P(p)$ and select the one with highest priority as $\Psi_q$ to fill first;
        \STATE{\quad\textbf{5.}} Select exemplars from observed region for $\Psi_q$ based on Eq. (\ref{simm});
        \STATE{\quad\textbf{6.}}
        Estimate the missing values in $\Psi_q$ by EPC in Eq. (\ref{epc}) or EPD in Eq. (\ref{epd});
        \STATE{\quad\textbf{7.}} Update $\Phi$ and $\Omega$;
        \STATE{\quad\textbf{8.}} Update the confidence $C(p)$ term in the priority as calculated in Eq. (\ref{pri});\\
        \textbf{end while}
        \STATE{\textbf{Output}:} Estimated radiomap $\Tilde{t}(\mathbf{Z}_p)$ of the protected area.
    \end{algorithmic}
\end{algorithm}
    \item \textbf{Exemplar Similarity}: After selecting patch $\Psi_q$ with the highest filling priority, we can find exemplars from observed regions $\Phi$ similar to $\Psi_q$ to assist missing value estimation. Since the template captures overall smooth patterns for all missing areas, we could utilize  landscape map $m(\mathbf{Z})$ and depth map $W(\mathbf{Z})$ to better select similar patches. More specifically,  distance between a candidate exemplar $\Psi_u$ and the target region $\Psi_q$ is calculated by
    \begin{align}\label{simm}
        \text{SIM}(\Psi_u, \Psi_q)=&a\cdot||(\Psi_q)_\Phi-(\Psi_u)_\Phi||_F^2+b\cdot||W(\Psi_q)-
        W(\Psi_u)||_F^2 \nonumber\\ 
        &+c\cdot||m(\Psi_q)-m(\Psi_u)||_F^2+d\cdot dis(q,u),
    \end{align}
    where the first term measures similarity of radio spectrum, the second term describes the similarity in depth map, the third term measures neighbor radio information, and the last term involves
distance between two patch locations. In general, we set larger values for $a$ and $b$ to 
allow higher
impact on the radio spectrum and depth map. From the exemplar patch, we can 
apply dictionary learning or direct copy as EPD/EPC, respectively, to fill the missing areas in $\Psi_q$.
\end{itemize}

\noindent
\underline{2-C. Combination of Template and Perturbation:} After inpainting the missing regions $\Tilde{h}(\mathbf{Z}_p)$ in perturbation and $\Tilde{t}(\mathbf{Z}_p)$ in the template, we obtain $\Tilde{r}(\mathbf{Z}_p)$ {according to Eq. (\ref{pert_con})} as
\begin{equation}
    \Tilde{r}(\mathbf{Z}_p)=\Tilde{t}(\mathbf{Z}_p)+\Tilde{h}(\mathbf{Z}_p).
\end{equation}
{We further smooth the estimated $\Tilde{r}(\mathbf{Z}_p)$ for position compensation}. Among various smoothing schemes, such
as inverse gradient weight smooth \cite{c48} or the $l_0$ gradient minimization \cite{c49},
we apply gradient inverse weighted smoothing for location correction. 

We summarize the algorithm of template inpainting as Algorithm \ref{alg:B}, and summmarize
the two-step template-perturbation radiomap reconstruction with depth map as Algorithm \ref{alg:C}.

\begin{algorithm}[t]
    \caption{Template-Perturbation Radiomap Inpainting (TPI)}
    \label{alg:C}
    \begin{algorithmic}
        \STATE {\textbf{Input}}: Radiomap $r(\mathbf{Z})\in\mathbb{R}^{P\times Q}$ and the restricted area $\mathbf{Z}_p$ with size $M\times N$ (observed region is denoted by $\Phi$ while missing region $\Omega$ is initialized by $\mathbf{Z}_p$), and urban landscape map $m(\mathbf{Z})\in\mathbb{R}^{P\times Q}$.
        \STATE{\textbf{1. }} Construct superpixels of the whole region $\mathbf{Z}\in\mathbb{R}^{P\times Q}$ from $m(\mathbf{Z})$ based on ERS as Eq. (\ref{obj}), and build the template radiomap $t(\mathbf{Z})\in\mathbb{R}^{P\times Q}$ based on Eq. (\ref{template_con});
         \STATE{\textbf{2. }} Construct the perturbation radiomap $h(\mathbf{Z})\in\mathbb{R}^{P\times Q}$ from Eq. (\ref{pert_con});
          \STATE{\textbf{3. }} Inpaint the perturbation radiomap as $\Tilde{h}(\mathbf{Z})$ based on Algorithm \ref{alg:A};
           \STATE{\textbf{4. }} Inpaint the template radiomap as $\Tilde{t}(\mathbf{Z})$ based on Algorithm \ref{alg:B};
           \STATE{\textbf{5. }} Combine the estimated template and perturbation for the estimated radiomap of the restricted areas $\mathbf{Z}_p$ by $\Tilde{r}(\mathbf{Z}_p)=\Tilde{t}(\mathbf{Z}_p)+\Tilde{h}(\mathbf{Z}_p)$;
        \STATE{\textbf{6. }} Apply the gradient inverse weighted smoothing on $\Tilde{r}(\mathbf{Z}_p)$ for location correction;
        \STATE{\textbf{Output}:} Estimated radiomap $\Tilde{r}(\mathbf{Z}_p)$ of the protected area.
    \end{algorithmic}
\end{algorithm}

\section{Experimental Results} \label{results}
We now present test results of the proposed radiomap inpainting algorithms. More specifically, we test the EPD/EPC with a small-scale high-fidelity JHU-APL (Johns Hopkins University Applied Physics Laboratory) dataset. We also test the template-perturbation TPI with a larger-scale BRATlab dataset.

\subsection{Small-Scale High-Resolution radiomap Inpainting}
We first present the results of EPD/EPC for high-resolution radiomap inpainting.

\subsubsection{Data Informtion and Preprocessing} 
\begin{figure}[t]
	\centering
	\subfigure[Mean Power (watt).]{
		\label{img1b}
	\includegraphics[height=3cm,width=4cm]{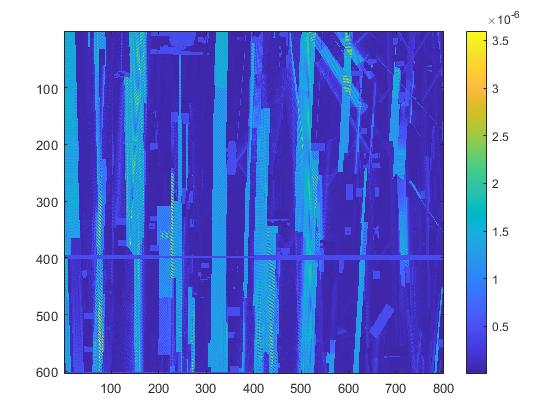}}
	\hspace{0.8cm}
	\subfigure[Satellite]{
		\label{img2b}
	\includegraphics[height=3cm,width=3cm]{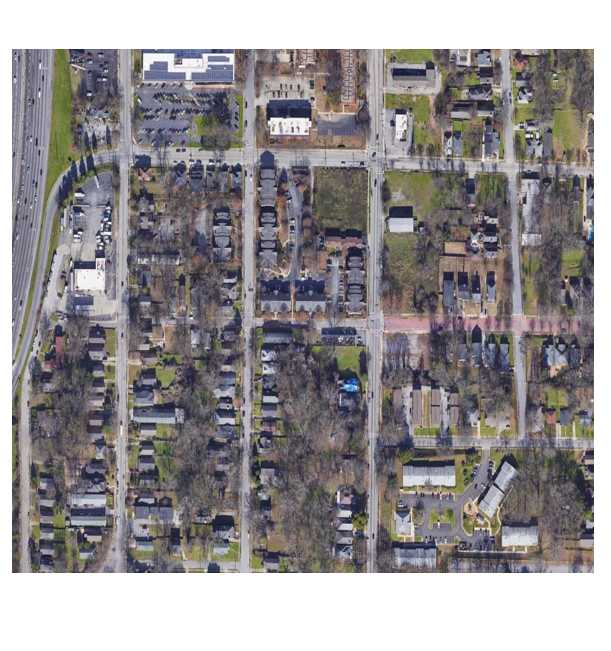}}
	\caption{Illustration of APL dataset.}
	\label{data_img}
\end{figure}

\begin{figure}[t]
	\centering
	\subfigure[]{
		\label{img11b}
		\includegraphics[height=3cm,width=4cm]{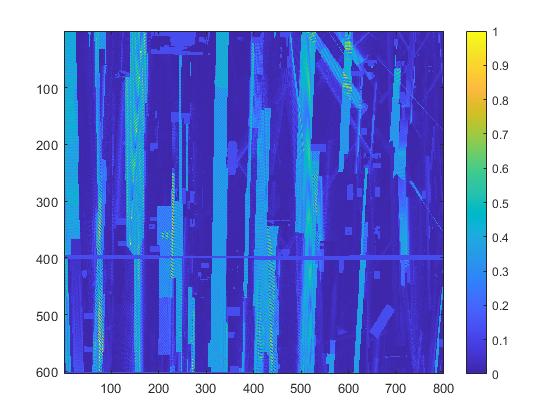}}
	\subfigure[]{
		\label{img22b}
		\includegraphics[height=3cm,width=4cm]{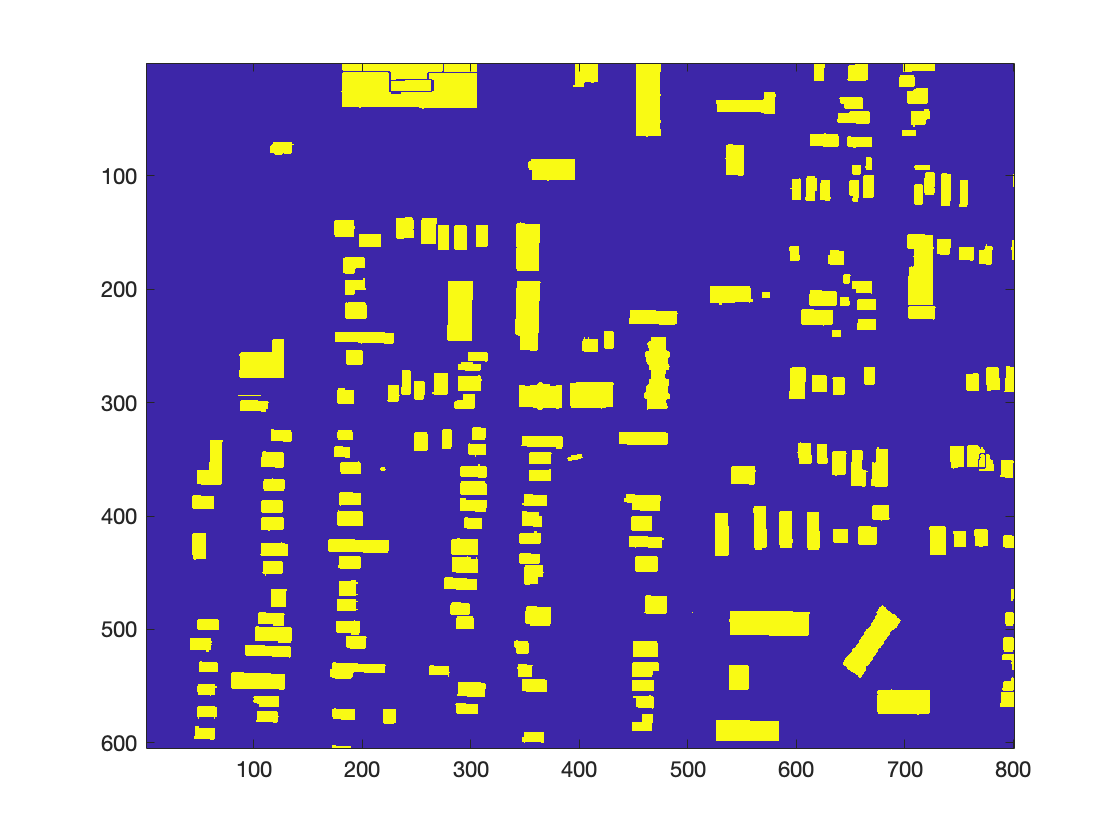}}
	\subfigure[]{
		\label{img33b}
		\includegraphics[height=3cm,width=4cm]{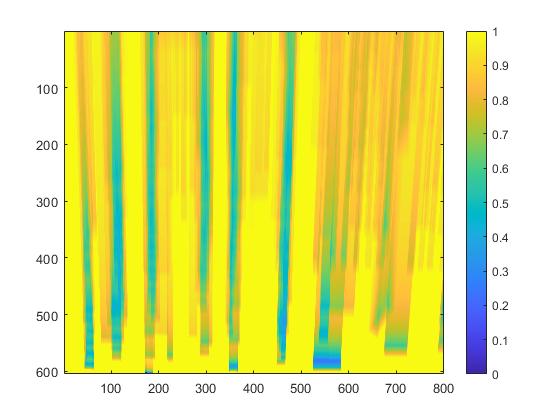}}
	\caption{Preprocessing of APL Dataset: a) Normalized radiomap; b) Segmented buildings; and c) Block term in priority.}
	\label{data_img1}
\end{figure}

We test the proposed methods in the Johns Hopkins University Applied Physics Laboratory (JHU-APL) dataset, reportedly generated from the \textit{Wireless inSite Software} \cite{c50} with Light Detection and Ranging (LIDAR) information. This APL dataset targets a designated region in Atlanta, Georgia, USA, and the resolution of the LIDAR data is 1-meter. The location of the area is at 33.7283$\sim$33.7327 in latitude and -84.3923$\sim$-84.3854 in longitude.
The APL dataset is for
one transmitter (TX) whose signal is received in a 10-block area.
The TX antenna is a uniform square array of $16\times 16$ elements, spaced at 0.5 wavelength. The TX is located at latitude/longitude of 33.689/-84.390. Its antenna height is 201 meters, with center frequency of 2660 MHz. The single antenna receivers are at
height of 2.01 meters and uniformly
spaced by 0.8 meters. We average the antenna gains from the TX for each location and conform it to a $604\times 800$ grid, i.e., $r(\mathbf{Z})\in\mathbb{R}^{604\times 800}$. The generated radiomap, together with its satellite image of nearby landscape, are displayed in Fig. \ref{data_img}. Here, we further normalize the radiomap measurements between 0$\sim$1 for convenience. Since the original radiomap and the normalized one have exactly the same patterns, we can apply
inpainting to the normalized radiomap and transform back without loss. To calculate the building block term in Eq. (\ref{btm}), we segment the buildings against the background from the satellite image. We show the normalized radiomap, segmented buildings, and $B(\mathbf{Z})$ in Fig. \ref{data_img1}.


\begin{figure}[t]
	\centering
	\subfigure[]{
		\label{ss1}
		\includegraphics[height=3cm,width=4cm]{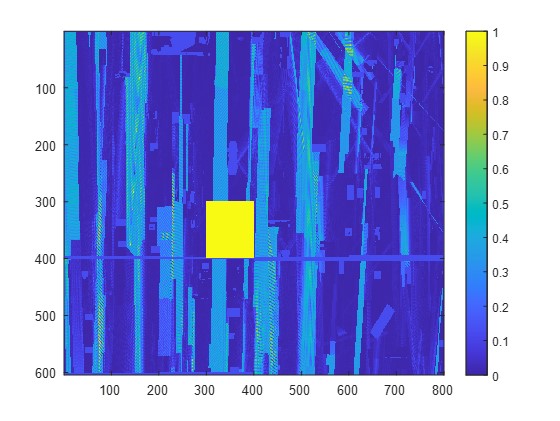}}
	\hspace{0.8cm}
	\subfigure[]{
		\label{ss2}
	\includegraphics[height=3cm,width=4cm]{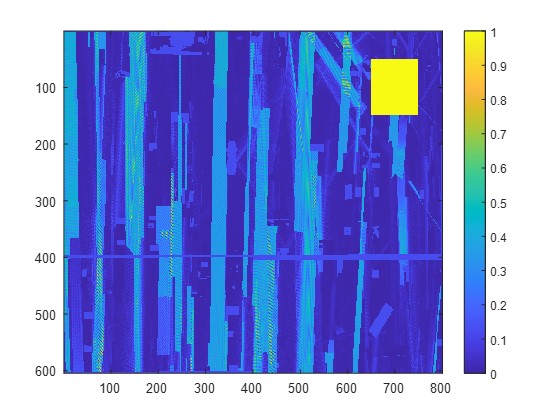}}
	\caption{Selected Areas to Test Performance: a) Scenario with smooth spectral pattern; and b) Scenario with complex neighborhood landscape. The restricted/inaccessible areas $\mathbf{Z}_p$ with size $100\times 100$ are marked in yellow.}
	\label{sss}
\end{figure}

\subsubsection{Performances in Different Scenarios} \label{sec: psa}
To evaluate the proposed methods, we first consider two specific scenarios: 1) one with smooth spectral pattern and regular neighborhood landscape shown as Fig. \ref{ss1}; and b) one with complex neighborhood landscape as shown in Fig. \ref{ss2}. In both scenarios, we consider a restricted area with size $100\times 100$, marked as yellow in Fig. \ref{sss}. The PSD in the whole restricted area is unavailable, which we need to reconstruct from the observed radiomaps in $r(\mathbf{Z})$.

Since the observations are limited for a single region and are insufficient for training DL neural networks, we mainly use non-DL approaches as benchmarks. More specifically, we compare our methods with model-based interpolation (MBI) \cite{c6}, RBF interpolation (RBF) \cite{c7}, exemplar-based inpainting (EI) \cite{c32}, dictionary learning (DL) \cite{c33}, and label propagation (LP) \cite{c51}. Here, MBI and RBF are model-based methods related to the distance from TX. EI and DL are image inpainting approaches without considering
radio propagation. For LP, we combine the satellite images and information on TX locations as features. In addition to proposed EPC/EPD, we also test texture priority $P(\cdot)D(\cdot)$ together with block term $B(\cdot)$ under exemplar-based copy (EBC). 
To allow fair comparisons, we select the patch size of $\Psi_p$ as $21\times 21$. For methods related to dictionary learning, we apply the K-SVD \cite{c43} for dictionary generation and set the size of dictionary as $K=500$.

\begin{figure*}[t]
	\centering
	\subfigure[Reconstructed radiomap for Scenario 1.]{
		\label{r1}
		\includegraphics[width=6.5in]{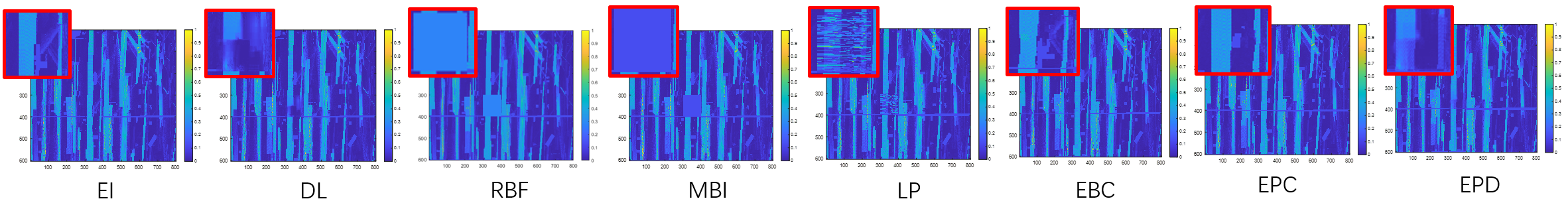}}\\
	\subfigure[Reconstructed radiomap for Scenario 2.]{
		\label{r2}
		\includegraphics[width=6.5in]{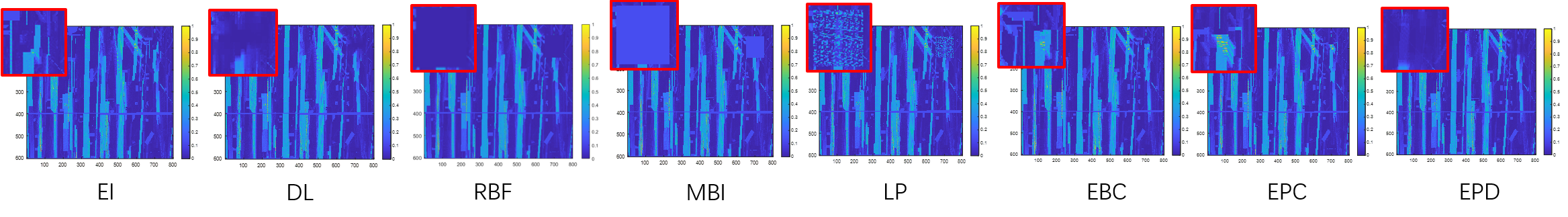}}
	\caption{Visualized Results in Selected Areas: (a) and (b) describe the regular and complex area, respectively; the results in red blocks are zoom-in presentations pf $\Tilde{r}(\mathbf{Z}_p)$.}
	\label{rr}
\end{figure*}
\begin{table*}[t]
	\caption{Numerical Results in Selected Areas.}
	\centering
	\begin{tabular}{|l|l|l|l|l|l|l|l|l|}
		\hline
		& EI     & DL     & RBF    & MBI    & LP     & EBC    & EPC             & EPD             \\ \hline
		MSE in Scenario 1 & 0.0092 & 0.0152 & 0.0448 & 0.0271 & 0.0327 & 0.0088 & \textbf{0.0038} & 0.0096          \\ \hline
		MSE in Scenario 2 & 0.0258 & 0.0158 & 0.0217 & 0.0173 & 0.0306 & 0.0227 & 0.0152          & \textbf{0.0136} \\ \hline
	\end{tabular}
\label{tt}
\end{table*}

To visualize, the reconstructed radiomaps are shown in Fig \ref{rr}, and the corresponding numerical results are shown as Table \ref{tt}. Here, we apply standard MSE defined by
\begin{equation}
    \mbox{MSE}=\frac{1}{m}\sum_{i=1}^m (x_i-\tilde{x}_i)^2,
\end{equation}
 where $\tilde {x}_i, i=1,\cdots,m$ are the estimated radiomap.
As shown in Fig. \ref{rr}, MBI is unable to 
accurately predict the radiomap for the missing area since PSD in this 
dataset is over smaller distance variation from the transmitter but is more sensitive
to the surrounding landscape as 
 seen from Fig. \ref{data_img}. The RBF interpolation also fails because of unevenly-distributed observed samples.
 The results of LP appear noisy since training features from environment are with low quality. The proposed methods based on radio propagation 
 priority exhibit superior performance compared with traditional image inpainting methods,
 benefiting from the enhanced features and textures based on propagation information.
 As shown in Fig. \ref{img33b}, the propagation priority terms favor the vertical direction to fill the region, which is consistent with the apparent spectrum pattern in Fig. \ref{img11b}. EPC displays sharper features while EPD provides more robust but more blurry results. 
 In the first scenario with smooth spectrum patterns, EPC displays significant improvement since the vertical patterns therein are clearer. In the second scenario near obstacles, EPC sometimes lead to unexpected artifacts while EPD 
 exhibits better robustness. The numerical results in Table \ref{tt} are consistent with the 
 visual results. Thus, one can choose whether EPC or EPD should be applied according to the complexity of the nearby landscape.

\begin{figure*}[t]
	\centering
	\subfigure[MSE.]{
		\label{mse}
		\includegraphics[height=4cm]{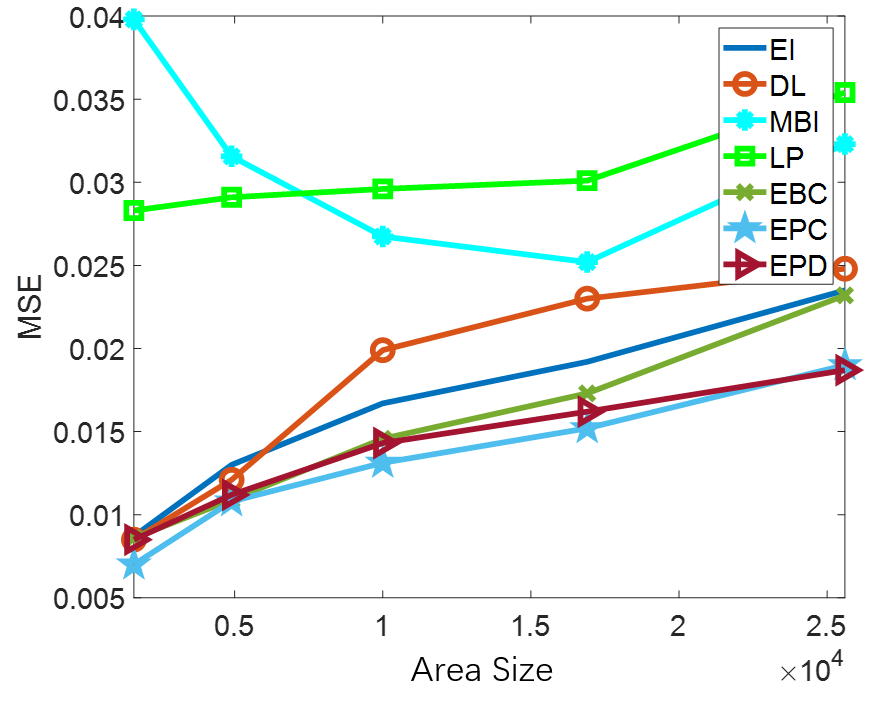}}
	\hspace{2cm}
	\subfigure[NE]{
		\label{ne}
		\includegraphics[height=4cm]{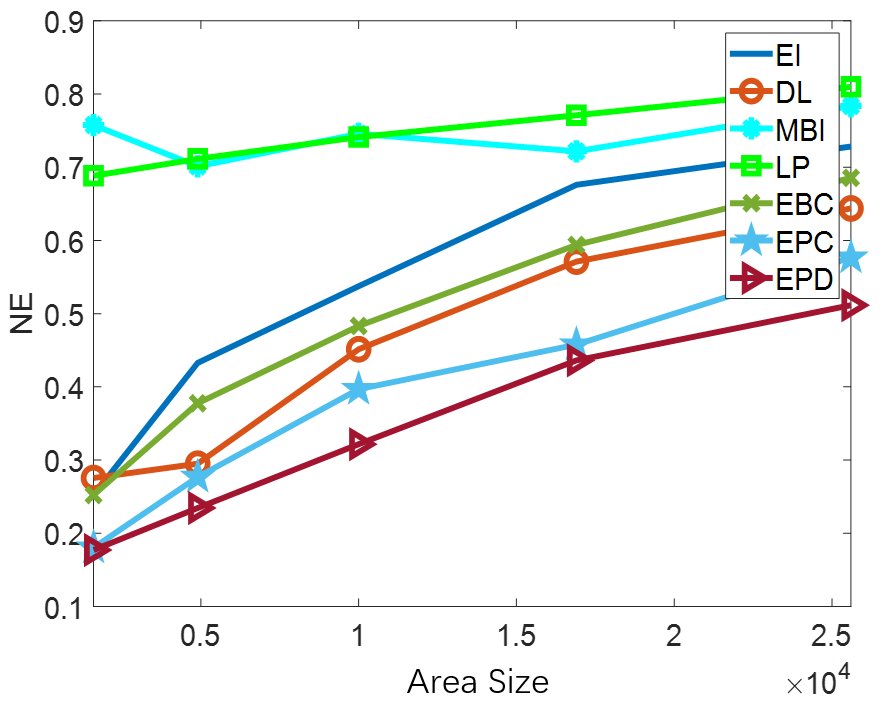}}
	\caption{Numerical Results in Different Area Sizes}
	\label{err}
\end{figure*}

 \subsubsection{Performances for Different Area Sizes}
To evaluate the performances in more general setups, we compare different methods for restricted areas with various area sizes, i.e., $30\times 30$, $70\times 70$, $100\times 100$, $130\times 130$, and $160\times 160$. We randomly generate 10  $\mathbf{Z}_p$ for each region size, and calculate the mean error of reconstructed radiomap for different generated areas. In addition to MSE, we also examine the performance via normalized error ({NE}) denoted by 
\begin{equation}
    \mbox{NE}=\frac{\sum_{i=1}^m(x_i-\tilde{x}_i)^2}{\sum_{i=1}^m x_i^2}.
\end{equation}
The results are shown in Table. \ref{err}. The estimation error increases with growing
restricted regions since neighbor spectrum information is more limited and less accurate for larger areas.
Since the APL dataset is more sensitive to the landscape rather than the Tx distance, MBI fails to capture clear spectral patterns and exhibits consistently poorer performance. 
Overally, our proposed methods achieve the best performance by comparison with the
list of tested methods.  EPC and EPD show similar MSE results while EPD generates better 
normalized erorr than EPC. 
The results indicate that EPC works better in some special scenarios whereas EPD is more robust. The conclusions are similar to Section \ref{sec: psa} and further demonstrate the benefits of the proposed method.

 \subsection{Large-Scale radiomap Reconstruction}
 We next present the results of large-scale radiomap reconstruction for
 restricted areas for the proposed EPD/EPC and TPI.

\begin{figure}[t]
	\centering
	\subfigure[City Map.]{
		\label{img1b1}
  \includegraphics[height=3cm,width=4cm]{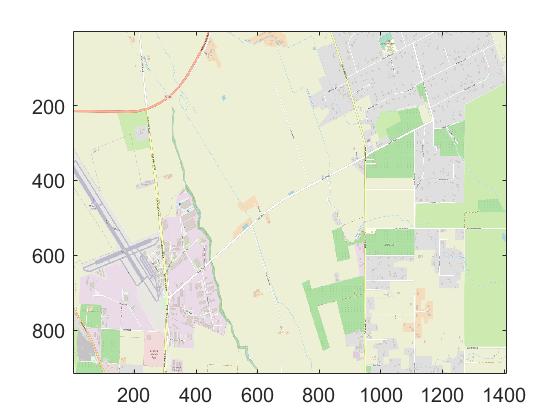}}
	\hspace{0.8cm}
	\subfigure[Building Segmentation]{
		\label{img2b1}
	\includegraphics[height=3cm,width=4cm]{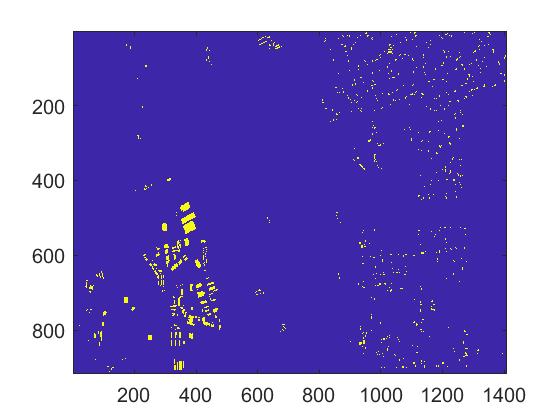}}
	\caption{landscape of Braltab Dataset: a) Nearby envrionment; b) Segmented buildings which are marked as yellow.}
	\label{data_img11}
\end{figure}
\begin{figure}[t]
	\centering
	\subfigure[TX Location.]{
		\label{imgc1}
  \includegraphics[height=3cm,width=3.5cm]{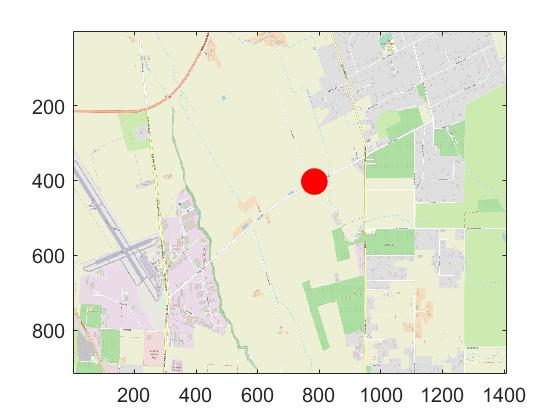}}
	\hspace{0.0cm}
	\subfigure[radiomap - 1 TX]{
		\label{imgc2}
\includegraphics[height=3cm,width=4cm]{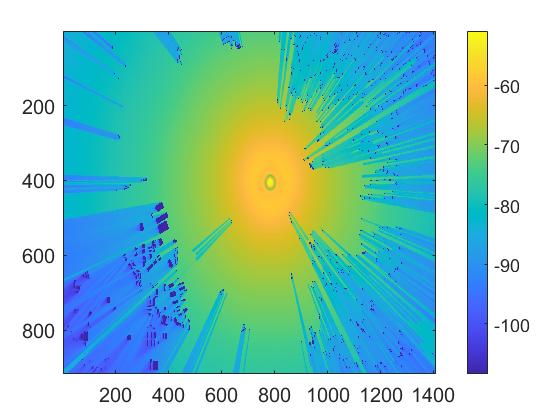}}
 \hspace{0.0cm}
	\subfigure[TX Locations.]{
		\label{imgc3}
	\includegraphics[height=3cm,width=3.5cm]{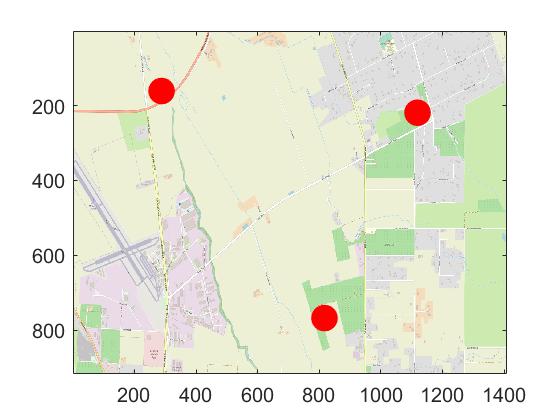}}
 \hspace{0.0cm}
	\subfigure[radiomap - 3 TX]{
		\label{imgc4}
	\includegraphics[height=3cm,width=4cm]{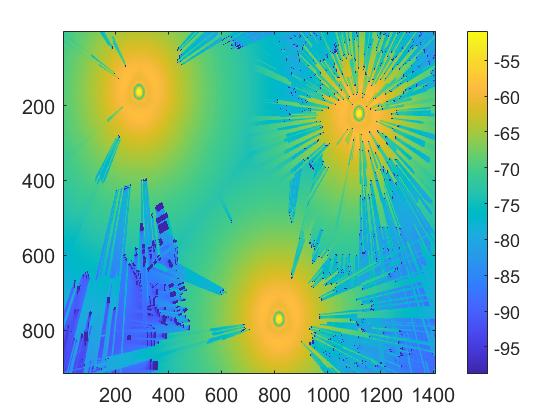}}
	\caption{Illustration of Bratlab Dataset: a) - b) TX locations and radiomap for the single-transmitter scenario; c) - d) TX locations and radiomap for the Triple-transmitters scenario. The TX locations are marked as red, and the radiomap is in dBm.}
	\label{data_imgc}
\end{figure}
 \subsubsection{Data Information and Preprocessing}
We evaluate the proposed algorithms in the BRATlab Dataset generated by the simulation tool Altair Feko \footnote{\url{https://www.altair.com/feko/}}. The dataset is generated for locations in
California, USA, for which landscape map is accessible from OpenStreetMap \footnote{\url{https://www.openstreetmap.org/#map=5/38.007/-95.844}}. The locations are
centered at $36.8767\sim36.9170$ in latitude and $-121.4176\sim-121.3375$ in longitude.
The whole region is conformed as a $917\times 1409$ regular grid.
There are 1120 buildings in this region and each building is set as 17.5 meters.
We segment the buildings from the background to calculate the depth map shown as Fig. \ref{data_img11}. 
To measure the performances of the proposed methods, we consider two scenarios: 1) 1-transmitter (TX) scenario; and 2) 3-TX scenario. For both scenarios, each transmitter contains three antennas (model 720842A2) with height as 35 meters and initial power as 46.00 dBm. The frequency used for each antenna is 2625 MHz. In the 3-TX scenario, the three TXs are located at $(161, 288)$, $(219, 1119)$, and $(768, 817)$ in the grid, respectively. In the 1-TX scenario, the TX is located as $(403, 784)$. The full radiomaps are shown in Fig. \ref{data_imgc}.

\begin{figure}[t]
	\centering
	\subfigure[Radio Depth Map.]{
		\label{depc1}
  \includegraphics[height=3cm,width=3.5cm]{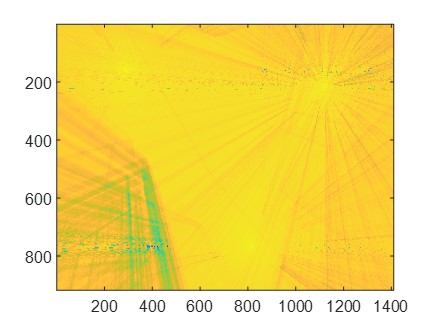}}
	\hspace{1cm}
	\subfigure[Scenario 1.]{
		\label{depc2}
\includegraphics[height=3cm,width=4cm]{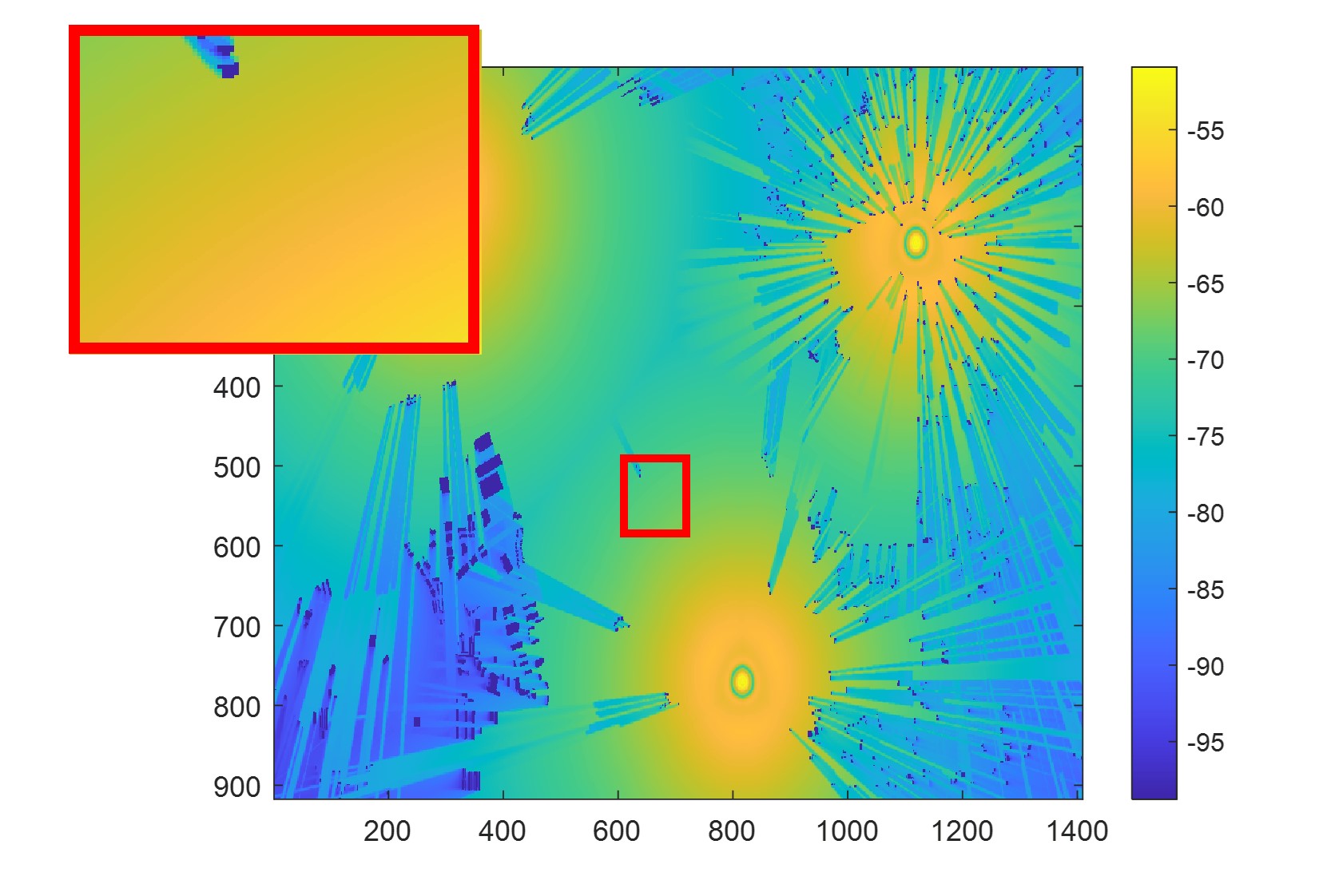}}
 \hspace{1cm}
	\subfigure[Scenario 2.]{
		\label{depc3}
	\includegraphics[height=3cm,width=4cm]{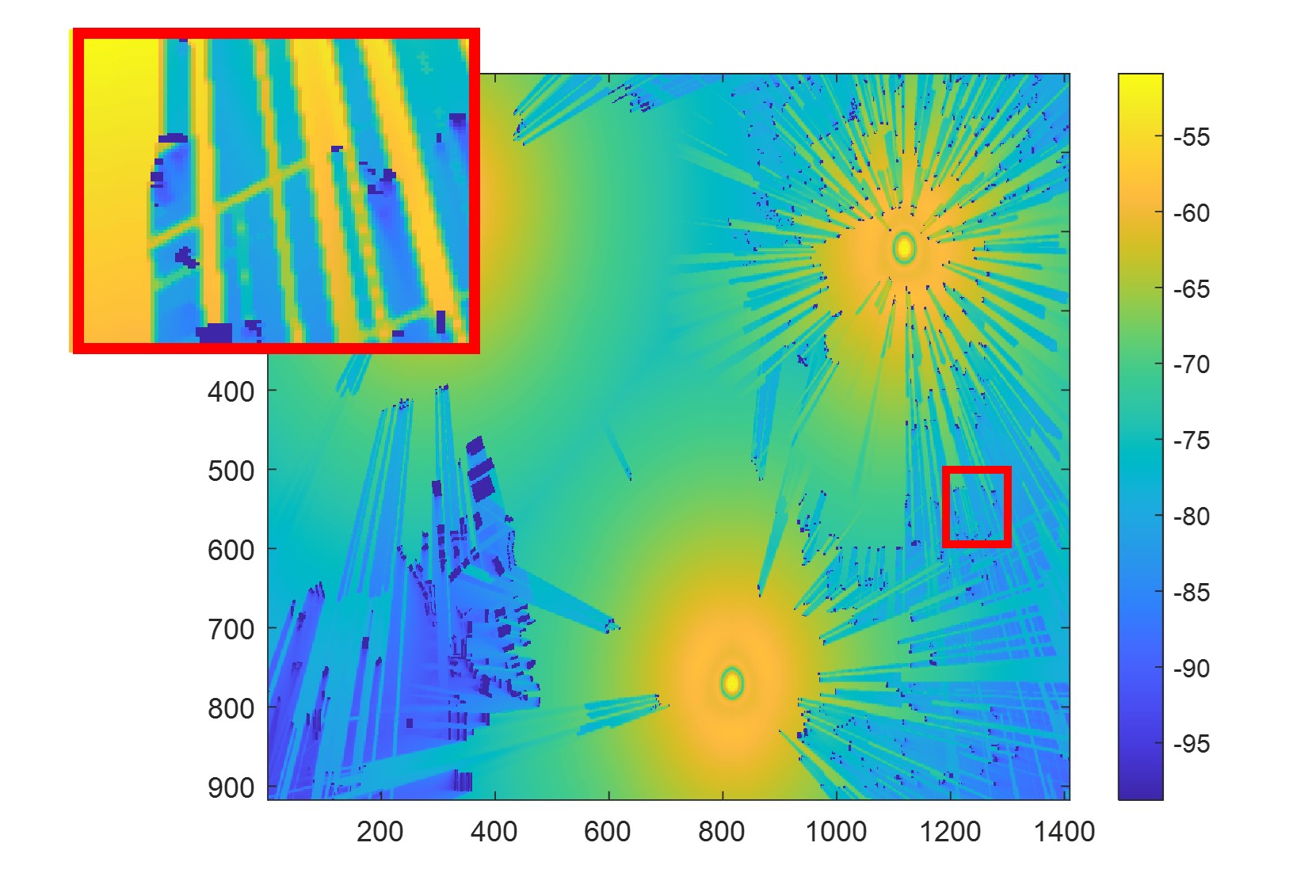}}
	\caption{Illustration of Scenarios: 1) Radio depth map for 3-TX Bratlab dataset; 2) Scenario 1 with regular patterns; and 3) Scenario 2 with complex patterns. The restricted area is zoomed in within the red block.}
	\label{depc}
\end{figure}

\begin{table}[t]
\centering
\caption{MSE of Reconstructed Restricted Areas for Different Methods in Bratlab Dataset}
\begin{tabular}{|l|l|l|l|l|l|l|l|l|l|}
\hline
           & MBI    & RBF    & EI     & DL     & NN     & LP     & EPC    & EPD    & TPI    \\ \hline
Scenario 1 & 18.47  & 16.37  & 18.40  & 18.78  & 12.29  & 20.41  & 14.57  & 16.40  & \textbf{8.37}   \\ \hline
Scenario 2 & 139.32 & 130.29 & 137.33 & 126.11 & 115.21 & 123.11 & 121.76 & 111.07 & \textbf{104.01} \\ \hline
\end{tabular}
\label{mse_tpi}
\end{table}

\begin{figure}[t]
    \centering	
     \includegraphics[height=6.3cm]{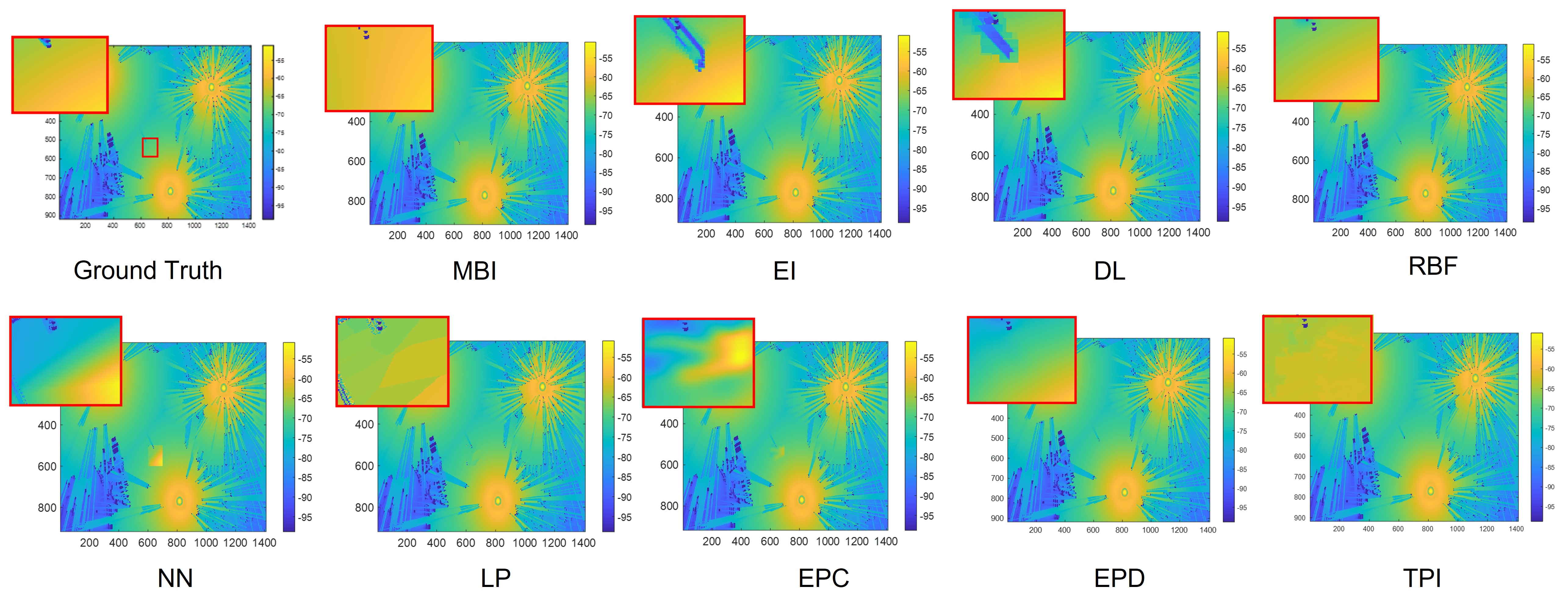}
     \caption{Visual Results of Reconstructed radiomap for Scenario 1.}
     \label{example:vrs1}
\end{figure}
\begin{figure}[t]
    \centering	
     \includegraphics[height=6.2cm]{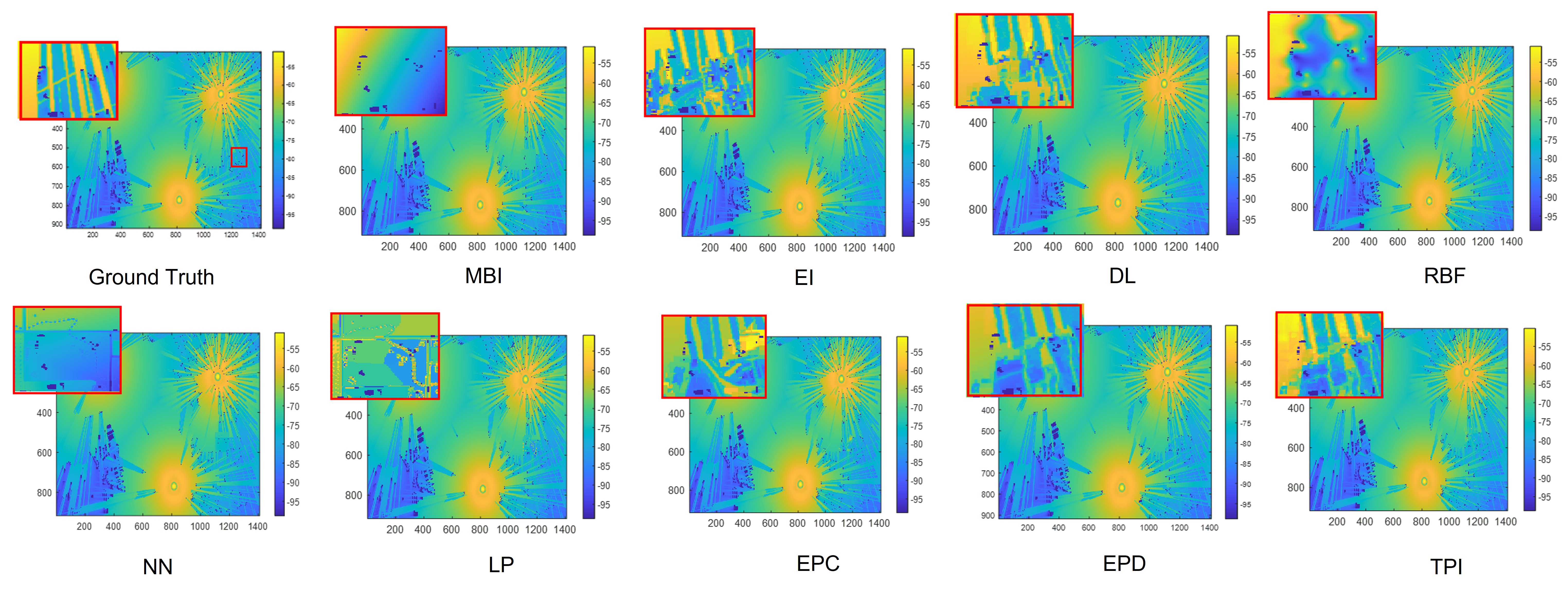}
     \caption{Visual Results of Reconstructed radiomap for Scenario 2.}
     \label{example:vrs2}
\end{figure}
\subsubsection{Performance in Selected Regions}
Similar to the small-scale radiomap reconstruction, we also first consider two specific scenarios: 1) one with regular patterns which has similar landscape to other observed regions; 2) one with complex nearby landscape, which has more unique spectral patterns in the radiomap. We select two $100\times 100$ regions as restricted areas in the 3-TX radiomaps, respectively, shown as Fig. \ref{depc2} and Fig. \ref{depc3}. We compare our methods with the MBI, RBF, and LP interpolations discussed in Section \ref{sec: psa}. We also present the comparison of our template-perturbation radiomap inpainting (TPI) with dictionary learning (DL), image inpainting (EI), deep regression neural networks (NN) and our proposed EPC/EPD. For learning-based methods, such as LP and NN, we input the distance to the TXs, the landscape map and segmented building images as features. For the proposed TPI, we apply Eq. (\ref{dep_idw}) for constructing radio depth map as shown in Fig. \ref{depc1}, and set the parameter $\sigma=0.01$. The patch size is set as $15\times 15$ for all inpainting-based approaches, and the dictionary size is $K=500$ for DL-related methods.

Table \ref{mse_tpi} provides the numerical MSE results, and Fig. \ref{example:vrs1} - \ref{example:vrs2}
provide the corresponding the image results for visualization.  From the numerical results, our proposed TPI achieves superior performance to all other tested approaches. Fig. \ref{example:vrs1}
shows that the model-based interpolation favors radio propagation model and generates  smoother patterns, which could better capture scenarios with regular spectral patterns. 
However, when dealing with more complex patterns like Scenario 2,  model-based interpolation
methods, such as MBI and LP, fail to capture the
smaller-scale shadowing effects. On the other hand, learning-based methods, including LP and NN, could 
provide good approximation on the overall PSD values if there are enough training samples. 
However, since the landscape map is rather noisy in real-life scenarios, 
learning-based approaches sometimes introduce unexpected artifacts, as shown in Fig. \ref{example:vrs2}. Compared with traditional image inpainting methods, such as EI and DL, our model-based inpainting significantly improve radiomap reconstruction, demonstrating the benefits of our proposed radio-based priority and radio depth map. To further evaluate the efficiency of the proposed radio depth map, we also evaluate the performance of TPI in randomly selected $100\times 100$ missing areas in the APL dataset. The overall MSE, as shown in Table \ref{tp1}, demonstrates better performance of the proposed EPC/EPD and TPI.
\begin{table}[t]
\centering
\caption{MSE in a Randomly Selected Region in APL Dataset}
\begin{tabular}{|l|l|l|l|l|l|l|l|l|}
\hline
Methods & MBI    & EI     & NN     & LP     & DL     & EPC    & EPD    & TPI    \\ \hline
MSE     & 0.0231 & 0.0153 & 0.0184 & 0.0317 & 0.0144 & 0.0118 & 0.0102 & \textbf{0.0086} \\ \hline
\end{tabular}
\label{tp1}
\end{table}

\subsubsection{Performance over Different Restricted Region Sizes} 
\begin{figure}[t]
    \centering	
     \includegraphics[height=6cm]{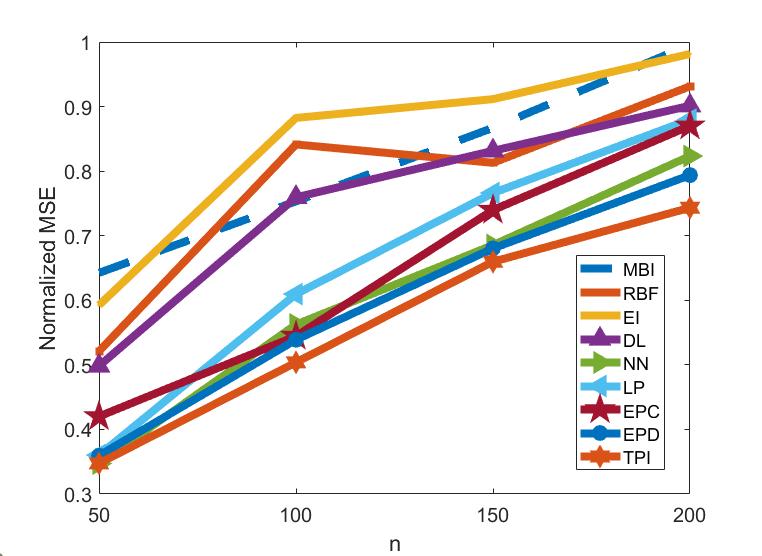}
     \caption{MSE over Different Sizes of Restricted Areas.}
     \label{example:mse}
\end{figure}

Suppose that each restricted area has size $n\times n$. We evaluate the performances in regions with different sizes $n$ ranging from $50 \sim 200$.
To evaluate the overall performance, we randomly select 20 regions from the 1-TX dataset and 3-TX dataset as the restricted areas and calculate the average MSE. We normalize the maximal MSE among all the methods as 1 to display a more general curve, shown in Fig. \ref{example:mse}. 
These results show TPI with the best performances and EPD/EPC with competitive results in comparison to DL-based approaches. The experimental results demonstrate the efficiency of the proposed methods and the defined radio depth map.

\subsubsection{Potential Integration of Proposed Depth Map and Neural Networks} 

\begin{figure}[t]
	\centering
	\subfigure[Ground Truth.]{
		\label{NNc1}
  \includegraphics[height=3cm,width=4cm]{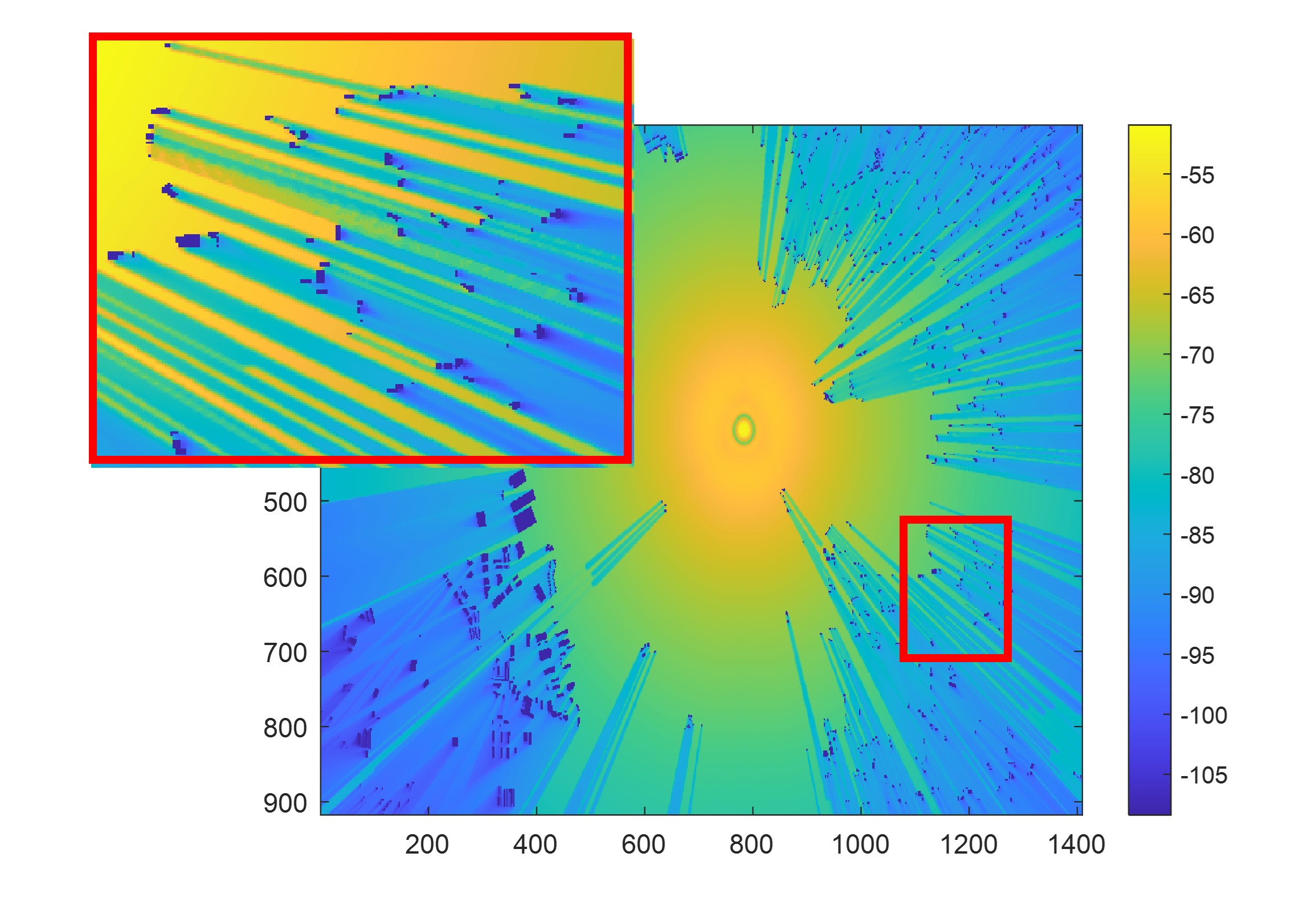}}
	\hspace{1cm}
	\subfigure[Landscape Map as Input: MSE=25.17.]{
		\label{NNc2}
\includegraphics[height=3cm,width=4cm]{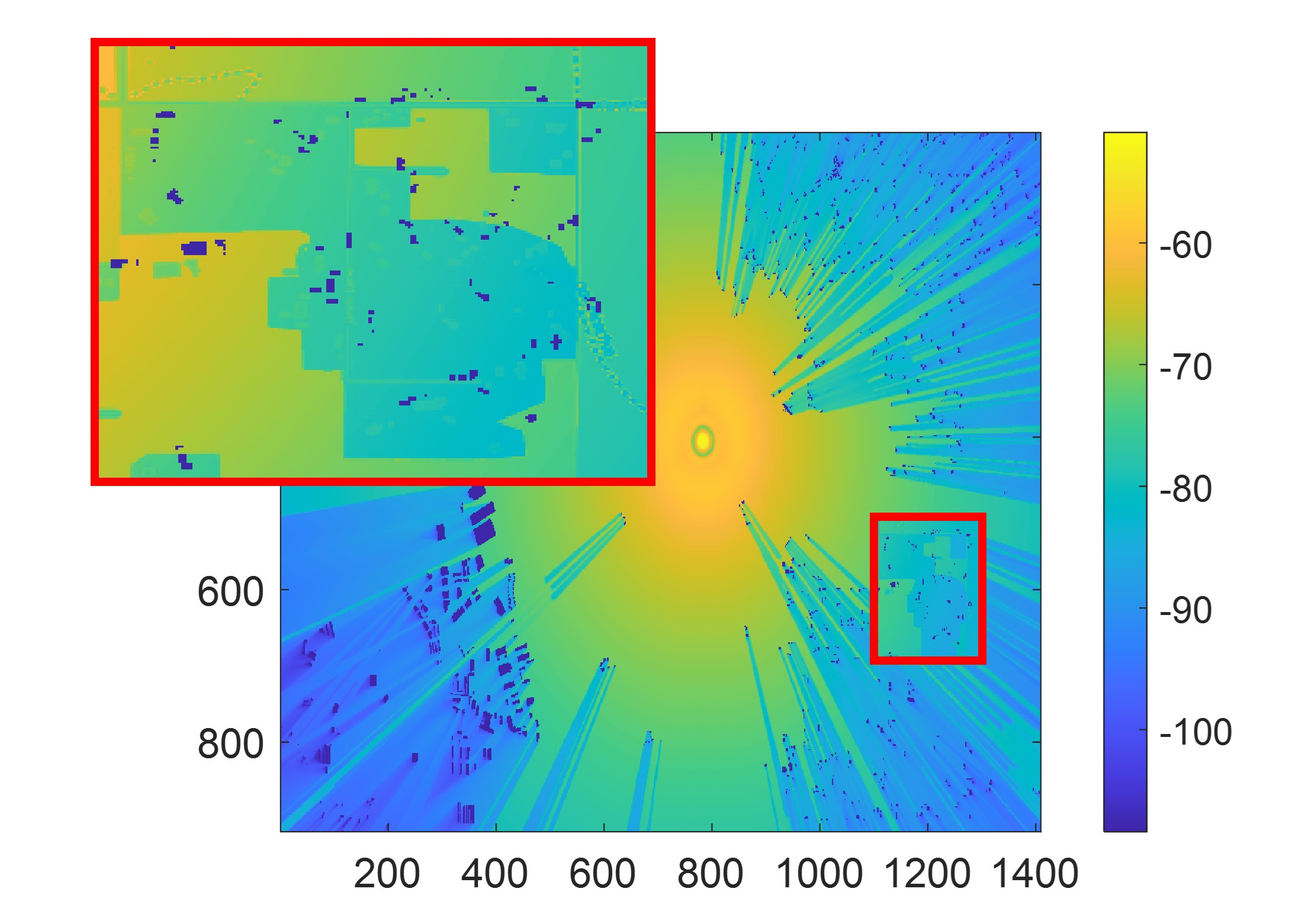}}
 \hspace{1cm}
	\subfigure[Radio Depth Map as Input:  MSE=8.74.]{
		\label{NNc3}
	\includegraphics[height=3cm,width=4cm]{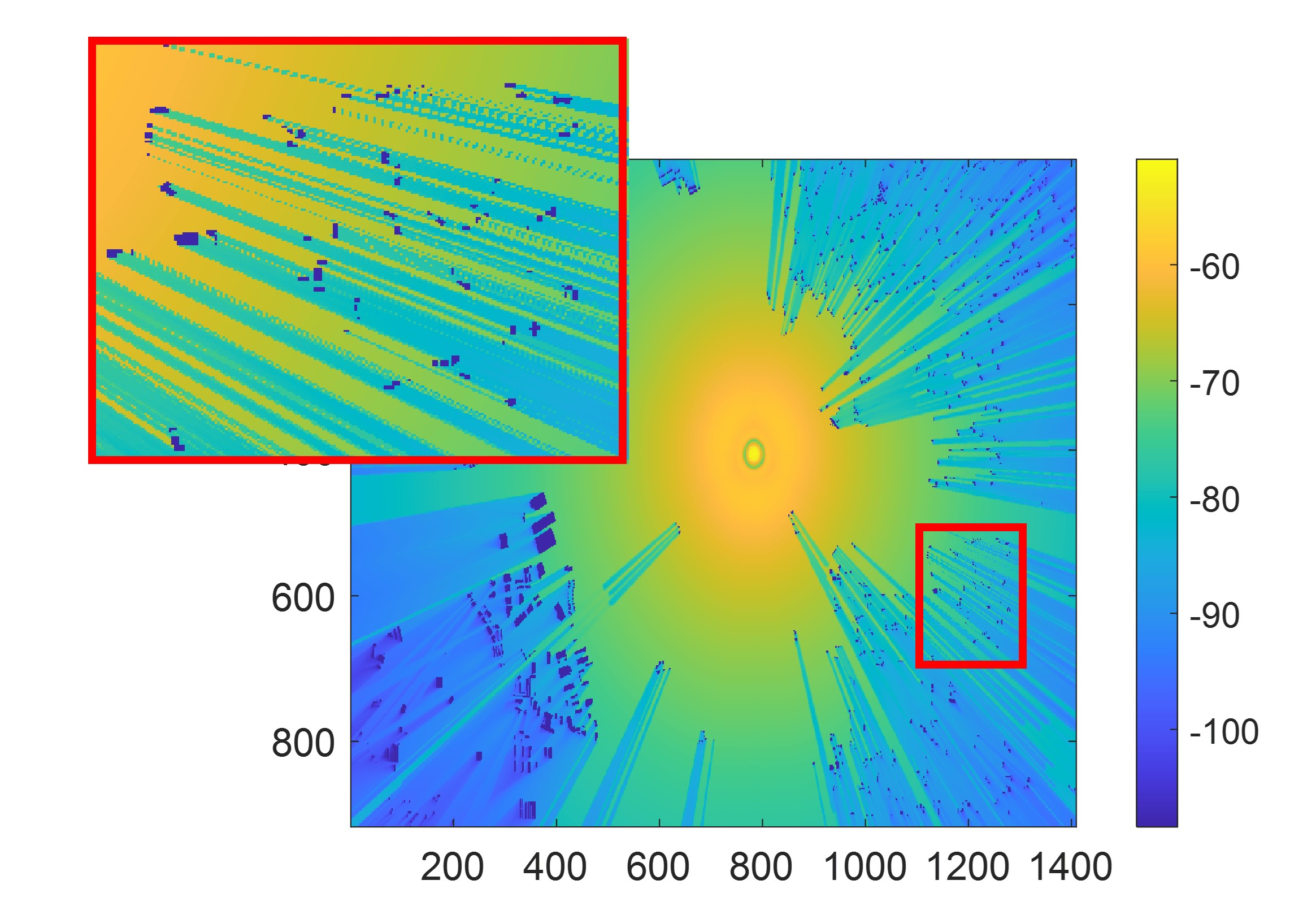}}
	\caption{Examples of the Results from Neural Networks with Different Input Features. }
	\label{NNc}
\end{figure}

Since landscape maps are often noisy and not accurate, treating landscape and TX locations as features may lead to unexpected artifacts in the radiomap reconstruction,
as shown in Fig. \ref{example:vrs2}. To further illustrate the potential power of the proposed radio depth map, we replace  landscape maps with radio depth map as the input feature for the DL neural networks. An example result is shown as Fig. \ref{NNc}, where the radio depth map leads to more accurate patterns and smaller MSE for DL neural networks, demonstrating the potential integration of the proposed radio depth map and DL in related future research. We plan  to explore more systematic ways to combine model-based radiomap inpainting with DL in  future works.

\section{Conclusion} \label{concolusion}
In this work, we propose two novel algorithms for radiomap inpainting for restricted regions. To integrate
well known radio propagation model with texture patterns, we first define a radio propagation priority and introduce an exemplar-based approach for small-scale fine-resolution radiomap inpainting. To generalize radiomap inpainting, we further define a new radio depth map and develop a two-step template-perturbation algorithms for large-scale radiomaps. We evaluate our proposed algorithms by using two different datasets. Our test results
demonstrate the efficiency of the propagation-based priority and the radio depth map to spatially capture radio PSD patterns. {The proposed methods can benefit spectrum access and management particularly for the restricted or sensitive areas that either
lack or has limited measurement access. }

With rapid growth and expansion of IoT and 5G systems, potential future directions should consider wireless network optimization and spectrum planning based on radiomap recovery and prediction. One potential direction is the timely fault detection of radio networks based on the sparse spectral measurement observations from sensors and mobile devices. Another promising direction may consider
integrating deep learning models into radiomap inpainting. {Beyond direct radiomaps  applications, our proposed radio depth map could efficiently model spectrum distribution and surrounding environments, which is expected to have broader potential impacts on wireless communication and IoT systems.}
We plan to explore these and other related directions in the future works.


 





\vfill


\begin{thebibliography}{1}
\bibliographystyle{IEEEtran}

\bibitem{c1} D. Romero and S. J. Kim, ``Radiomap estimation: a data-driven approach to spectrum cartography," in \textit{IEEE Signal Processing Magazine}, vol. 39, no. 6, pp. 53-72, Nov. 2022.

\bibitem{c2} S. Bi, J. Lyu, Z. Ding and R. Zhang, ``Engineering radiomaps for wireless resource management," in \textit{IEEE Wireless Communications}, vol. 26, no. 2, pp. 133-141, April 2019.

\bibitem{c3} M. S. Riaz, H. N. Qureshi, U. Masood, A. Rizwan, A. Abu-Dayya and A. Imran, ``Deep learning-based framework for multi-Fault diagnosis in self-healing cellular networks," 2022 \textit{IEEE Wireless Communications and Networking Conference (WCNC)}, Austin, TX, USA, 2022, pp. 746-751.

\bibitem{c4} S. Zhang and R. Zhang, ``Radiomap-based 3d path planning for cellular-connected uav," in \textit{IEEE Transactions on Wireless Communications}, vol. 20, no. 3, pp. 1975-1989, March 2021.

\bibitem{c5} Q. Luo, Y. Cao, J. Liu and A. Benslimane, ``Localization and navigation in autonomous driving: threats and countermeasures," in \textit{IEEE Wireless Communications}, vol. 26, no. 4, pp. 38-45, August 2019.

\bibitem{c6} M. Lee and D. Han, ``Voronoi tessellation based interpolation method for wi-fi radiomap construction,” in \textit{IEEE Communications Letters}, vol. 16, no. 3, pp. 404-407, Mar. 2012.

\bibitem{c7} A. Bazerque, G. Mateos, and G. B. Giannakis, ``Group-lasso
on splines for spectrum cartography," \textit{IEEE Trans. Signal Processing}, vol. 59, no. 10, Oct. 2011, pp. 4648–63.

\bibitem{c8} S. Kuo and Y. Tseng, ``Discriminant minimization search for large-scale rf-based localization systems," in \textit{IEEE Transactions on Mobile Computing}, vol. 10, no. 2, pp. 291-304, Feb. 2011.

\bibitem{c9} J. Krumm and J. Platt, ``Minimizing calibration effort for an indoor 802.11 device location measurement system," \textit{Microsoft Research}, Nov. 2003.

\bibitem{c10} R. Levie, C. Yapar, G. Kutyniok and G. Caire, ``Radiounet: fast radiomap estimation with convolutional neural networks," in \textit{IEEE Transactions on
Wireless Communications}, vol. 20, no. 6, pp. 4001-4015, Jun. 2021.

\bibitem{c11} S. Zhang, A. Wijesinghe, and Z. Ding. ``RME-GAN: A Learning Framework for Radio Map Estimation based on Conditional Generative Adversarial Network," in \textit{IEEE Internet of Things Journal}, May 2023.

\bibitem{c12} S. Zhang, T. Yu, J. Tivald, B. Choi, F. Ouyang and Z. Ding, ``Exemplar-Based Radio Map Reconstruction of Missing Areas Using Propagation Priority," \textit{GLOBECOM 2022 - 2022 IEEE Global Communications Conference}, Rio de Janeiro, Brazil, 2022, pp. 1217-1222.

\bibitem{c13}  A. E. C. Redondi, ``Radiomap interpolation using graph signal processing," in \textit{IEEE Comm. Letters}, vol. 22, no. 1, pp. 153-156, Jan. 2018.

\bibitem{c14} D. Schäufele, R. L. G. Cavalcante and S. Stanczak, ``Tensor completion for radiomap reconstruction using low rank and smoothness," \textit{2019 IEEE 20th International Workshop on Signal Processing Advances in Wireless Communications (SPAWC)}, Cannes, France, 2019, pp. 1-5.

\bibitem{c15} A. Criminisi, P. Perez and K. Toyama, ``Region filling and object removal by exemplar-based image inpainting," in \textit{IEEE Transactions on Image Processing}, vol. 13, no. 9, pp. 1200-1212, Sept. 2004.

\bibitem{c16} X. Fu, N. D. Sidiropoulos, J. H. Tranter and W. K. Ma, ``A factor analysis
framework for power spectra separation and multiple emitter localization,"
in \textit{IEEE Transactions on Signal Processing}, vol. 63, no. 24, pp. 6581-
6594, Dec.15, 2015.

\bibitem{c17} H. Braham, S. B. Jemaa, G. Fort, E. Moulines and B. Sayrac, ``Fixed
rank kriging for cellular coverage analysis," in \textit{IEEE Transactions on
Vehicular Technology}, vol. 66, no. 5, pp. 4212-4222, May 2017.

\bibitem{c18} J. Krumm and J. Platt, ``Minimizing calibration effort for an indoor 802.11
device location measurement system," \textit{Microsoft Research}, Nov. 2003.


\bibitem{c20} A. Khalajmehrabadi, N. Gatsis and D. Akopian, ``Structured group sparsity: a novel indoor wlan localization, outlier detection, and radiomap interpolation scheme," in \textit{IEEE Transactions on Vehicular Technology}, vol. 66, no. 7, pp. 6498-6510, July 2017.

\bibitem{c21} H. Sun and J. Chen, ``Propagation map reconstruction via interpolation assisted matrix completion," in \textit{IEEE Transactions on Signal Processing}, vol. 70, pp. 6154-6169, 2022.

\bibitem{c22} G. Boccolini, G. Hernandez-Penaloza and B. Beferull-Lozano, ``Wireless ˜
sensor network for Spectrum Cartography based on Kriging interpolation," \textit{2012 IEEE 23rd International Symposium on Personal, Indoor and Mobile Radio Communications - (PIMRC)}, Sydney, NSW, Australia, Sep.
2012, pp. 1565-1570.

\bibitem{c23} Y. Teganya and D. Romero, ``Deep completion autoencoders for radio
map estimation," in \textit{IEEE Transactions on Wireless Communications}, vol.
21, no. 3, pp. 1710-1724, Mar. 2022.

\bibitem{c24} C. Parera, Q. Liao, I. Malanchini, C. Tatino, A. E. C. Redondi and M.
Cesana, ``Transfer learning for tilt-dependent radiomap prediction," in
\textit{IEEE Transactions on Cognitive Communications and Networking}, vol.
6, no. 2, pp. 829-843, Jun. 2020.

\bibitem{c25} S. K. Vankayala, S. Kumar, I. Roy, D. Thirumulanathan, S. Yoon and I.
S. Kanakaraj, ``Radiomap estimation using a generative adversarial network and related business aspects," \textit{2021 24th International Symposium
on Wireless Personal Multimedia Communications (WPMC)}, Okayama,
Japan, Feb. 2021, pp. 1-6.

\bibitem{c26} C. H. Liu, H. Chang, and T. Park, ``Da-cgan: a framework for indoor radio design using a dimension-aware conditional generative adversarial network", \textit{Proceedings of the IEEE/CVF
Conference on Computer Vision and Pattern Recognition (CVPR) Workshops}, 2020, pp. 498-499.

\bibitem{c27} H. Zou et al., ``Adversarial learning-enabled automatic wifi indoor radio
map construction and adaptation with mobile robot,” in \textit{IEEE Internet of
Things Journal}, vol. 7, no. 8, pp. 6946-6954, Aug. 2020.


\bibitem{c29} Y. Zeng, X. Xu, S. Jin and R. Zhang, ``Simultaneous navigation and radio
mapping for cellular-connected uav with deep reinforcement learning," in
\textit{IEEE Transactions on Wireless Communications}, vol. 20, no. 7, pp. 4205-
4220, Jul. 2021.

\bibitem{c30} H. J. Bae and L. Choi, ``Large-scale indoor positioning using geomagnetic field with deep neural networks," \textit{2019 IEEE International
Conference on Communications (ICC)}, Shanghai, China, May 2019, pp. 1-6.

\bibitem{c31} O. Elharrouss, N. Almaadeed, S., Al-Maadeed, and Y. Akbari, ``Image inpainting: a review," \textit{Neural Processing Letters}, no. 51, pp. 2007-2028, Dec. 2019.

\bibitem{c32} A. Criminisi, P. Perez and K. Toyama, ``Region filling and object removal by exemplar-based image inpainting," in \textit{IEEE Transactions on Image Processing}, vol. 13, no. 9, pp. 1200-1212, Sept. 2004.

\bibitem{c33} Bin Shen, Wei Hu, Y. Zhang and Y. J. Zhang, "Image inpainting via sparse representation," \textit{2009 IEEE International Conference on Acoustics, Speech and Signal Processing}, Taipei, 2009, pp. 697-700.

\bibitem{c34} H. Mobahi, S. R. Rao and Yi Ma, "Data-driven image completion by image patch subspaces," \textit{2009 Picture Coding Symposium}, Chicago, IL, USA, 2009, pp. 1-4.

\bibitem{c35} H. Li, W. Luo and J. Huang, ``Localization of diffusion-based inpainting in digital images," in \textit{IEEE Transactions on Information Forensics and Security}, vol. 12, no. 12, pp. 3050-3064, Dec. 2017.

\bibitem{c36} C. S. Weerasekera, T. Dharmasiri, R. Garg, T. Drummond and I. Reid, ``Just-in-time reconstruction: inpainting sparse maps Using single view depth predictors as priors," \textit{2018 IEEE International Conference on Robotics and Automation (ICRA)}, Brisbane, QLD, Australia, 2018, pp. 4977-4984.

\bibitem{c37} Y. L. Chang, Z. Y. Liu and W. Hsu, ``Vornet: spatio-temporally consistent video inpainting for object removal," \textit{2019 IEEE/CVF Conference on Computer Vision and Pattern Recognition Workshops (CVPRW)}, Long Beach, CA, USA, 2019, pp. 1785-1794.

\bibitem{c38} Y. Zeng, J. Fu, H. Chao and B. Guo, ``Learning pyramid-context encoder network for high-quality image inpainting," \textit{2019 IEEE/CVF Conference on Computer Vision and Pattern Recognition (CVPR)}, Long Beach, CA, USA, 2019, pp. 1486-1494.

\bibitem{c39} H. Liu, B. Jiang, Y. Xiao and C. Yang, ``Coherent semantic attention for image inpainting," \textit{2019 IEEE/CVF International Conference on Computer Vision (ICCV)}, Seoul, Korea (South), 2019, pp. 4169-4178.

\bibitem{c40} Y. G. Shin, M. C. Sagong, Y. J. Yeo, S. W. Kim and S. J. Ko, ``Pepsi++: fast and lightweight network for image inpainting," in \textit{IEEE Transactions on Neural Networks and Learning Systems}, vol. 32, no. 1, pp. 252-265, Jan. 2021.

\bibitem{c41} J. Dong, R. Yin, X. Sun, Q. Li, Y. Yang and X. Qin, ``Inpainting of remote sensing sst images with deep convolutional generative adversarial network," in \textit{IEEE Geoscience and Remote Sensing Letters}, vol. 16, no. 2, pp. 173-177, Feb. 2019.

\bibitem{c42} N. M. Salem, H. M. K. Mahdi and H. Abbas, "Semantic image inpainting using self-learning encoder-decoder and adversarial loss," \textit{2018 13th International Conference on Computer Engineering and Systems (ICCES)}, Cairo, Egypt, 2018, pp. 103-108.


\bibitem{c43} M. Aharon, M. Elad, and A.M. Bruckstein, ``K-svd: design of dictionaries for sparse representation”, \textit{Proceedings of SPARSE}, Rennes, France, Nov. 2005, pp.9-12.

\bibitem{c44} S. G. Mallat and Zhifeng Zhang, ``Matching pursuits with time-frequency dictionaries," in \textit{IEEE Transactions on Signal Processing}, vol. 41, no. 12, pp. 3397-3415, Dec. 1993.

\bibitem{c45} S. Zhang, Q. Deng, and Z. Ding, ``Multilayer graph spectral analysis for hyperspectral images," \textit{EURASIP Journal on Advances in Signal Processing}, vol. 1, p. 92, Oct. 2022.

\bibitem{c46} M. Y. Liu, O. Tuzel, S. Ramalingam, and R. Chellappa, ``Entropy rate superpixel segmentation," in \textit{IEEE CVPR}, Colorado Springs, CO, USA, Jun. 2011, pp. 2097-2104.

\bibitem{c47} F. Qi, J. Han, P. Wang, G. Shi, and F. Li, ``Structure guided fusion for depth map inpainting," \textit{Pattern Recognition Letters}, vol. 34, no. 1, pp. 70-76, Jan. 2013.

\bibitem{c48} D.C. Wang, A.H. Vagnucci, and C.C. Li, ``Gradient inverse weighted smoothing scheme and the evaluation of its performance," \textit{Computer Graphics and image processing}, vol. 15, no. 2, pp. 167-181, Feb. 1981.

\bibitem{c49} L. Xu, C. Lu, Y. Xu, and J. Jia, ``Image smoothing via L0 gradient minimization," in \textit{Proceedings of the 2011 SIGGRAPH Asia conference}, New York, NY, USA, Dec. 2011, pp. 1-12.

\bibitem{c50} P. Mededovic, M. Veletic and Z. Blagojevic, ``Wireless insite software verification via analysis and comparison of simulation and measurement results," in \textit{2012 Proceedings of the 35th International Convention MIPRO}, Opatija, Jul. 2012, pp. 776-781.

\bibitem{c51} Y. Ni, J. Chai, Y. Wang, and W. Fang, ``A fast radiomap construction method merging self-adaptive local linear embedding (lle) and graph-based label propagation in WLAN fingerprint localization systems," \textit{Sensors}, vol. 20, no. 3, p. 767, Jan. 2020.



\end{thebibliography}
\end{document}